\documentclass[conference]{IEEEtran}
\usepackage{cite}
\usepackage{amsmath,amssymb,amsfonts}
\usepackage{algorithmic}
\usepackage{graphicx}
\usepackage{textcomp}
\usepackage{fancyhdr}
\usepackage[hyphens]{url}
\newcommand{\ignore}[1]{}
\usepackage[pass]{geometry}
\usepackage{fancyhdr}
\usepackage[normalem]{ulem}
\usepackage[hyphens]{url}
\usepackage{hyperref}
\usepackage{color}
\usepackage{soul}
\usepackage{subcaption}
\usepackage{enumitem}
\usepackage[table,xcdraw]{xcolor}
\usepackage{soul}

\def\BibTeX{{\rm B\kern-.05em{\sc i\kern-.025em b}\kern-.08em
    T\kern-.1667em\lower.7ex\hbox{E}\kern-.125emX}}

\title{Understanding Training Efficiency of Deep Learning Recommendation Models at Scale}

\author{Bilge Acun, Matthew Murphy, Xiaodong Wang, Jade Nie, Carole-Jean Wu, Kim Hazelwood \\
\\
Facebook\\
acun@fb.com}

\IEEEoverridecommandlockouts\IEEEpubid{\makebox[\columnwidth]{* To appear in HPCA 2021.\hfill}
\hspace{\columnsep}\makebox[\columnwidth]{ }}

\begin{document}
\maketitle
\pagestyle{plain}

\begin{abstract}
The use of GPUs has proliferated for machine learning workflows and is now considered mainstream for many deep learning models. Meanwhile, when training state-of-the-art personal recommendation models, which consume the highest number of compute cycles at our large-scale datacenters, the use of GPUs came with various challenges due
to having both compute-intensive and memory-intensive components.
GPU performance and efficiency of these recommendation models are largely affected by model architecture configurations such as dense and sparse features, MLP dimensions. Furthermore, these models often contain large embedding tables that do not fit into limited GPU memory. The goal of this paper is to explain the intricacies of using GPUs for training recommendation models, factors affecting hardware efficiency at scale, and learnings from a new scale-up GPU server design, Zion.
\end{abstract}

\section{Introduction}
Deep learning recommendation algorithms have recently gained significant 
adoption to power a wide variety of products. For example, Google
utilizes deep candidate generation and ranking models to produce video suggestions 
in YouTube~\cite{Covington:RecSys2016}, and designs deep learning 
recommendation models to personalize mobile app suggestions for 
Google Play~\cite{Cheng:dlrs2016}.
Microsoft leverages deep learning recommendation models for news 
suggestions~\cite{Elkahky:www2015}, Netflix for personalized 
entertainment~\cite{raimond:2018}, Alibaba for product 
recommendations~\cite{zhou2019deep}. 
At Facebook, personalization and ranking use cases are powered 
by deep learning recommendation models: Instagram stories are 
ranked with multi-stage deep neural networks (DNNs)~\cite{Medvedev:2019}; 
Newsfeed Ranking and Search tasks are also built upon 
DNNs~\cite{hazelwood:hpca2018,Song:arxiv2020,naumov2019deep,gupta2020architectural,gupta2020deeprecsys}. 
Over the  last 18-month period,
we have witnessed the compute capacity for 
recommendation model training quadrupling at Facebook's datacenter 
fleet~\cite{naumov:arxiv2020}. Among the total AI training cycles 
at Facebook, more than 50\% has been devoted to training deep learning 
recommendation models.

As described in a recent work~\cite{hazelwood:hpca2018}, while language and 
vision models are trained on 8-GPU systems, 
the majority of deep learning recommendation models are trained on CPU servers. This is because of the large 
memory capacity and bandwidth requirement of embedding tables in these models.
The memory capacity of embedding tables have increased dramatically 
from tens of GBs to TBs throughout the industry~\cite{zhao2020distributed,yi:sysml2018,zhou2017deep,zhou2019deep,lui2020understanding}.
At the same time, memory bandwidth usage also increased quickly with the increasing number of embedding tables and the associated lookups.


\begin{figure}[t]
\centering
\includegraphics[width=0.9\columnwidth]{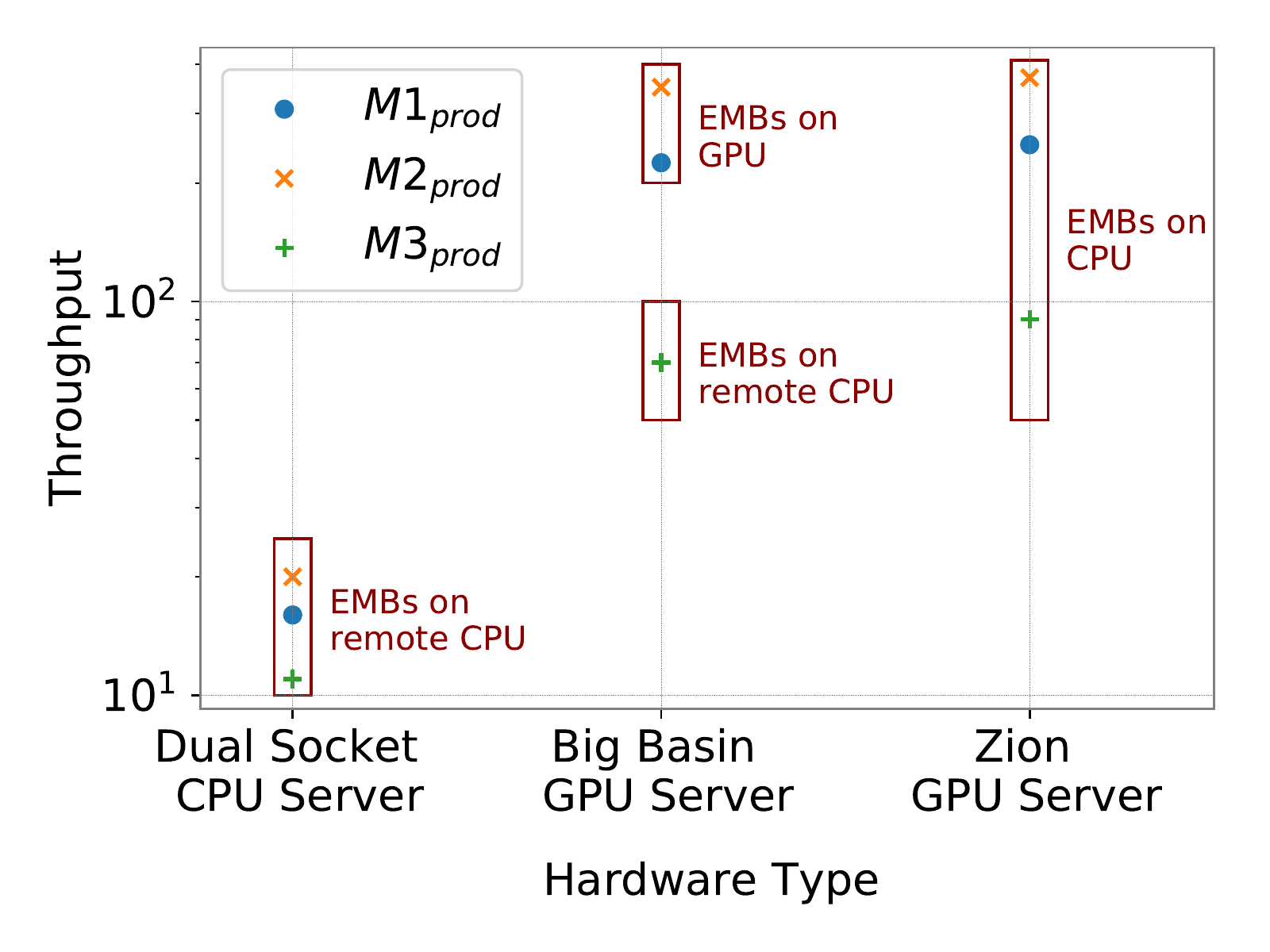}
\vspace{-0.3cm}
\caption{Throughput of three production models with different hardware and embedding table (EMB) placement strategies.}
\vspace{-0.5cm}
\label{speedup_quality_example}
\end{figure}

In order to leverage data and model-level parallelism in GPU 
training~\cite{nvidia_merlin,nvidia_mlperf}, we have 
devoted significant development efforts at Facebook to enable
production-scale recommendation models to train on
existing 8-GPU training systems (Big Basin).
At the same time, we started to prototype 
Facebook's next generation training systems (Zion)~\cite{zion}.
We enabled high performance training on these systems
by designing and implementing different placement
strategies for embedding tables.
We then characterized the interplay between model parameter
configurations and the underlying 
training system architectures to develop a better understanding
of GPU-training performance.

Figure~\ref{speedup_quality_example} depicts the relative throughput of
three different production recommendation models ($M1_{prod}$, $M2_{prod}$, $M3_{prod}$)
trained on Facebook's systems.
Overall, the training throughput increases as we go from
dual-socket CPU servers to Big Basin GPU server to Zion GPU server.
The degree of throughput improvement varies depending on the specific 
model parameters. For example, M3\textsubscript{prod} shows weaker performance scaling as
compared to M1\textsubscript{prod} and M2\textsubscript{prod} because of the significantly higher memory 
requirement of its embedding tables. Furthermore, placement strategies 
for embedding tables can influence the model training throughput 
significantly. For example, on the Big Basin GPU server, M1\textsubscript{prod} and M2\textsubscript{prod} 
training achieves the highest throughput when embedding tables are 
hosted on the GPU memory. Whereas for M3\textsubscript{prod}, the optimal embedding 
placement strategy shifts to the remote CPU, as tables do not fit
on the GPU memory of a single Big Basin server. On Zion GPU server,
optimal embedding placement is on system memory due to its large memory
capacity and high memory bandwidth.

Despite having similar overall structure and components, 
recommendation models can have diverse characteristics based on model 
architecture and input configurations. This can cause large variations 
in system resource utilization across training runs with different
configurations.
By analyzing over five hundred training runs, we show that there is a large variety of recommendation 
models being trained in Facebook datacenters, leading 
to different levels of utilization in CPUs, memory capacity, memory
bandwidth and network bandwidth.
Moreover, the most efficient choice of a hardware system depends
on the model parameters such as number of dense and sparse features,
embedding table sizes, feature interaction types, dimensions of
the multi-layer perceptron (MLP) stack. 

To select the optimal hardware system in a heterogeneous data-center
with a mix of CPU and GPU servers, 
there are methods to predict the performance of code,
such as predicting GPU performance using CPU 
code~\cite{ardalani2015cross}, using a roofline 
model~\cite{williams2009roofline}, or using a binary predictor
approach~\cite{baldini2014predicting}. However, such methods may not 
be directly applicable to recommendation models, since the
large memory capacity requirement of embedding tables requires different 
software infrastructure to handle the placement of the
embedding tables due to the limited memory of GPUs.

Performance characterization results presented 
in this paper sheds light on the effects of model
architecture configurations of deep learning recommendation models on 
training efficiency at scale, pinpointing system performance 
optimization opportunities.
Key contributions of this paper are as follows:

\begin{itemize}[leftmargin=*]
    \item We present the massive parameter design space for training production-scale deep learning recommendation models. The large variety of training experiments in the production environment contributes to 
    a wide range of CPU, memory, and networking utilization levels, 
    exposing ample performance optimization opportunities. 
    
    \item We pinpoint the development and system design challenges faced when training embedding tables using the CPU, Big Basin, and Zion training platforms. 
    \item In addition to the large memory 
    capacity demand for embedding tables, training throughput 
    can become limited by the often irregular vector accesses. 
    This is the first time detailed embedding table characterization 
    results on industry-scale recommendation model training is presented.
    
    \item Training efficiency, i.e. throughput per watt, on Big Basin improves by 4.3x, 2.8x for M1\textsubscript{prod} and M2\textsubscript{prod} respectively as compared to the baseline production CPU systems. Meanwhile, embedding-dominant recommendation models such as M3\textsubscript{prod} (with large embedding tables and more frequent embedding vector lookups) scales poorly on Facebook's Big Basin training systems.
    
    \item  When embedding tables do not fit onto a single Big Basin, training efficiency scaling of Facebook's Zion shines -- with its 2 TB of system memory and 1 TB/s memory bandwidth -- for embedding acceleration. As model sizes continue to grow into multiple TBs, it calls for novel solutions in order to achieve better performance and efficiency scaling across the entire system and infrastructure stack -- compute, networking, and storage.
 
\end{itemize}

\begin{figure}[b]
\centering
\includegraphics[width=0.7\columnwidth]{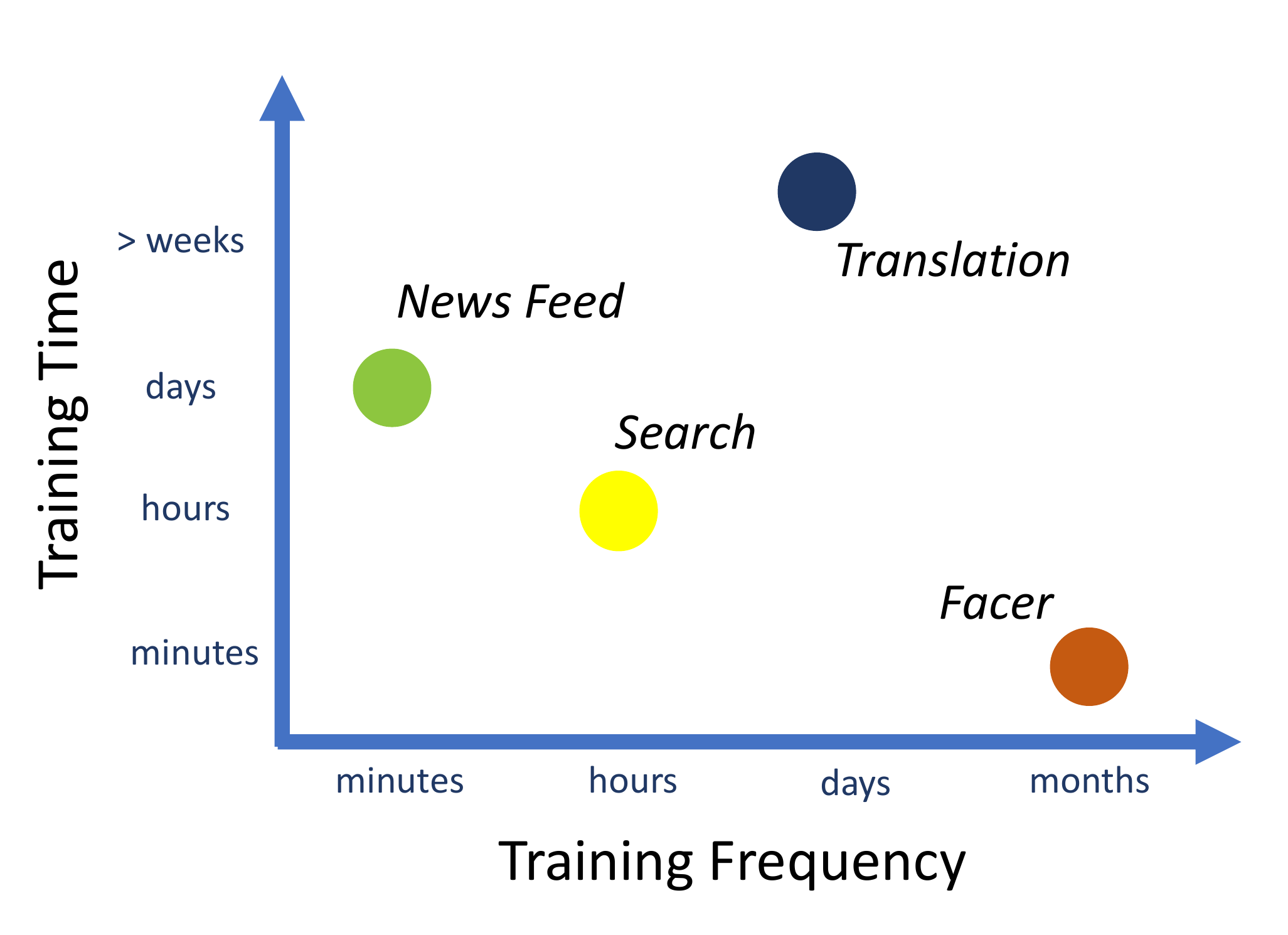}
\caption{Frequency and duration of various machine learning training workloads at Facebook.}
\label{fig:training_characteristics}
\end{figure}



\section{Background on ML Training at Facebook}
\label{background}

\begin{figure*}[t]
\centering
\includegraphics[width=\textwidth]{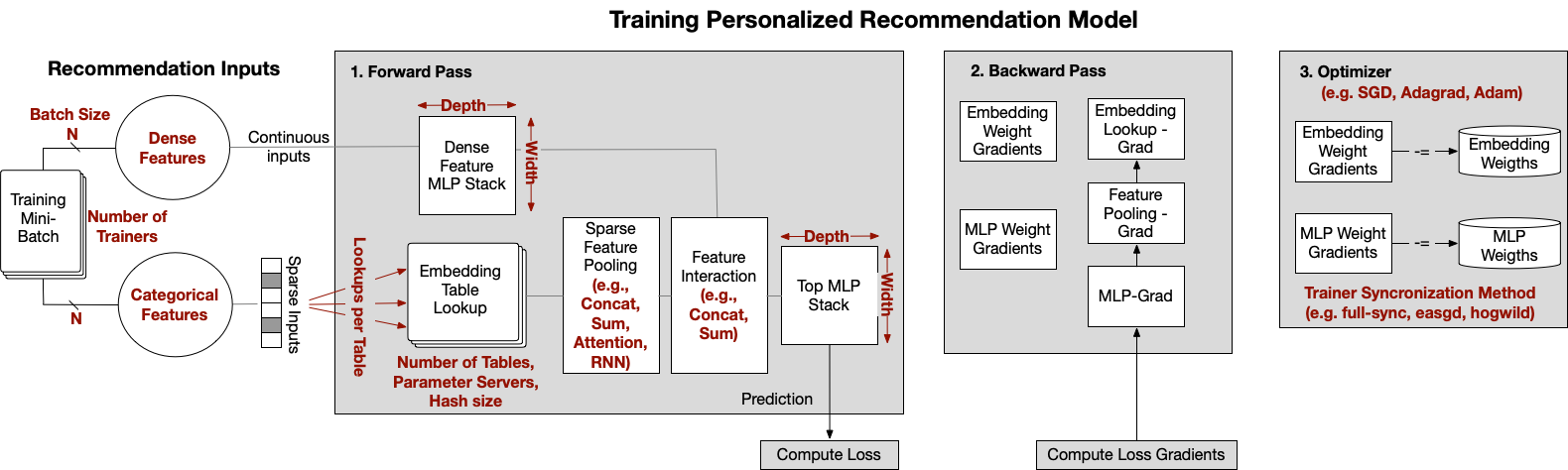}
\caption{General architecture of training personalized recommendation
models. Red components highlight different model configurations that affect efficiency.}
\vspace{-0.5cm}
\label{training_configs}
\end{figure*}

\subsection{Machine Learning Training at Facebook}

There are three primary execution phases for Facebook's machine learning (ML) pipeline: First, in the data pre-processing phase, we take unstructured data from persistent storage and manipulate it, in order to feed into a machine learning model. Second is the training phase, where we use important features from the data manipulation phase to build and train models. As the third and final step, the trained model is deployed for making real-time predictions in the inference phase~\cite{hazelwood:hpca2018}.

{\bf Not All Training Workflows Are Equal.} Depending on the workload, model training duration and training frequencies vary. Figure~\ref{fig:training_characteristics} depicts the training duration and frequency characteristics for various workloads: News Feed, Search, Language Translation, and Facer (for face detection).
Among the use cases, deep learning recommendation models,
i.e. News Feed and Search ranking~\cite{hazelwood:hpca2018,Song:arxiv2020}, are the most
frequently-trained models
while language translation uses recurrent neural network (RNN) variants~\cite{ott2019fairseq,arm2018loss,edunov2018understanding}, and image classification and object detection/tracking use convolutional neural network (CNN) variants~\cite{johnson2017billionscale,He_2017_ICCV,wu2018fbnet,lin2017focal}.
Furthermore, over the past eighteen months, the number of training workflow runs for deep learning recommendation models have
increased by 7 times. It is evident that {\it deep learning recommendation is one of the fastest growing training workloads at Facebook, demanding custom-built training systems to deliver high computation performance and optimized performance-cost efficiency.}

\subsection{Training Systems for Deep Learning Recommendation}

{\bf Recommendation Model Architectures.}
Figure~\ref{training_configs} illustrates the high-level architecture of training deep learning recommendation models. The two primary components are the Multi Layer Perceptron (MLP) layer modules and the Embedding Tables (EMB), which are used to learn latent space representations of dense and sparse information, respectively. 
The MLP layers are used to process continuous features while the EMBs process categorical features by transforming sparse, high-dimensional inputs to dense vectors. The configurations of these model components highly affect training
throughput and efficiency. We further elaborate on the components of
model training in Section~\ref{model_configs}~\&~\ref{hardware}.

\begin{figure}[t]
\centering
\includegraphics[width=0.9\columnwidth]{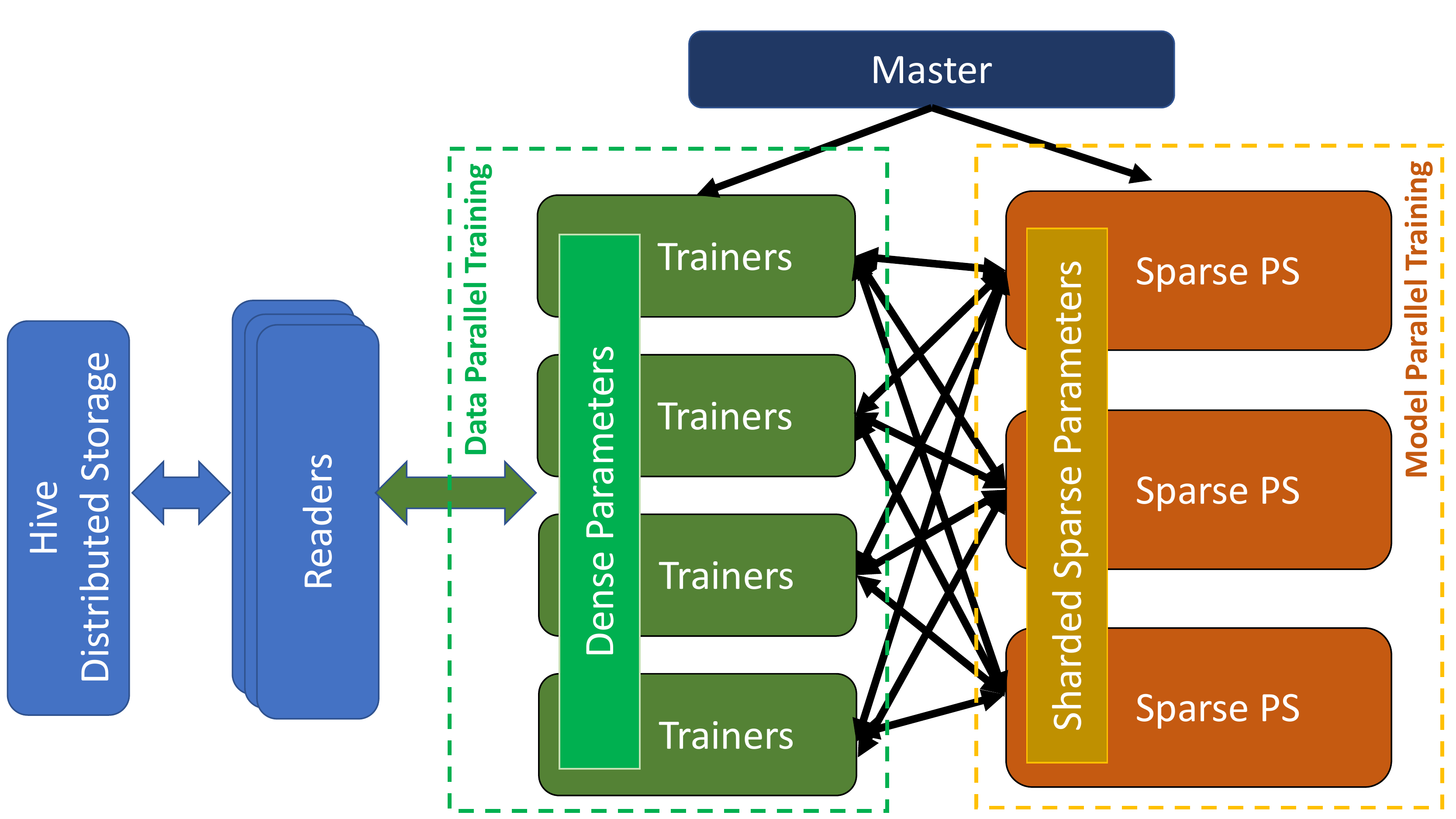}
\caption{Facebook's ML training pipeline for deep learning recommendation models.}
\vspace{-0.5cm}
\label{fig:training_system_overview}
\end{figure}

{\bf Distributed Training System Overview.}
Ever-increasing sizes of deep learning recommendation models, particularly embedding tables, introduce significant system design challenges.
This large memory capacity requirement, coupled with high 
throughput training performance demand, motivate at-scale distributed 
training system solutions. Figure~\ref{fig:training_system_overview} gives an overview for the distributed training system used for deep learning recommendation models in Facebook's production environment on CPU servers.



\textit{Training recommendation models exhibit both data parallelism and model parallelism.}
To exploit data parallelism, in production, each trainer server holds a copy of the MLP model parameters, read its own mini-batches from reader
servers and perform Elastic-Averaging SGD (EASGD)~\cite{zhang2015deep} update with the 
center, dense parameter server. Within a trainer, HogWild!~\cite{NIPS2011_4390} threads perform asynchronous updates on the model parameters.
As aforementioned, the large memory capacity requirement of embedding tables prevents direct model parameter replication. Thus, for embedding table training, we exploit model parallelism by partitioning embedding tables across multiple nodes, referred as sparse parameter servers.


\begin{figure*}[t!]
\centering
\begin{subfigure}[t]{0.49\textwidth}\centering
\includegraphics[width=\columnwidth]{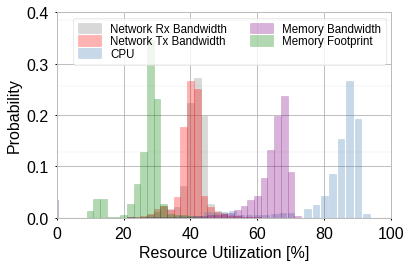}
\caption{Trainer Servers}
\end{subfigure}
\begin{subfigure}[t]{0.49\textwidth}\centering
\includegraphics[width=\columnwidth]{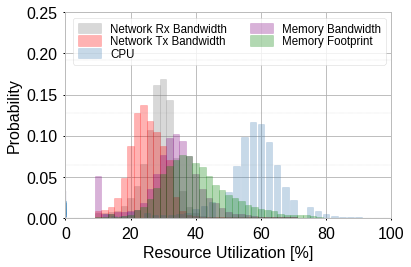}
\caption{Parameter Servers}
\end{subfigure}
\caption{Utilization distribution of a ranking model on a fixed scale (same number of trainer and parameter servers). Figure shows there is significant resource utilization variability from run to run training of a model type. The variability could be due to different model configurations and due to other system level variability.}
\vspace{-0.3cm}
\label{util_dist}
\end{figure*}

\textit{While training deep learning recommendation models on CPUs offer memory capacity advantage, a large degree of parallelism in the training process, which could be unlocked by utilizing accelerators, is left unexploited}. An alternative is to train recommendation models on Facebook's Big Basin GPU servers, originally designed for non-recommendation AI workloads~\cite{big_basin}.
Big Basin architecture features eight NVIDIA Tesla GPUs,
similar to NVIDIA's DGX-1 system~\cite{dgx}. As presented in Section~\ref{hardware}, training Facebook-scale recommendation models on the Big Basin system is not straightforward -- {\it optimization techniques tailored for the placement of embedding tables are needed to overcome the capacity requirement while addressing the potential increase in the embedding vector access latency.}



\section{Overview of Training Recommendation Model Architectures and Parameters}
\label{model_configs}

Diverse configurations of recommendation models affect the hardware
resource utilization at scale.
Utilization distribution of the CPU, memory and network resources resemble a wide Gaussian distribution which
we may partly attribute to different model configurations.
Figure~\ref{util_dist} shows this distribution when
running different instances of a particular model 
training over a week period at Facebook datacenters.
In this figure, we only include workflows that have the same model
type and system configurations i.e. same number of servers and hardware type.
Overall, trainer servers have high CPU and memory bandwidth utilization with relatively small variation. 
On the other hand, utilization distribution
is wider for parameter servers with a lower mean value and
a longer tail.
This variability could be partly due to system or hardware level variability~\cite{dean2013tail},
however wider distribution of the parameter servers
implies that the model architecture configuration does
have a significant effect on the hardware resource utilization.

ML engineers set up and tune different configurations of a model
through an internal recommendation model framework built for ease
of experimentation.
In this section, we give an overview commonly configured model
architecture components that affect hardware efficiency.
Some hyper-parameters, such as learning-rate, number of warm-up
iterations and optimizer algorithm have significant
effect on model quality while having minor or no effect on
performance. Therefore, we exclude those parameters in this paper.

\subsection{Model Architecture Configurations}

\subsubsection{\textbf{Feature Selection}} Feature selection is the 
process of selecting a subset of the most useful features to use in a 
specific ranking model. There are thousands of features to choose from
as an input, which are categorized into two distinct types: dense and 
sparse. Dense features are scalar inputs, whereas sparse features 
often encode categorical traits or relevant IDs.

\textbf{Dense Features.} In a typical recommendation model, 
pre-processed dense features are concatenated to a dense feature 
vector and passed through a dense architecture, such as a sequence of 
linear layers and activations. Computational cost of each dense 
feature is roughly the same. 

\textbf{Sparse Features.} For each sparse feature $X_i$, we map all possible indices of $X_i$ to a randomly-initialized embedding of dimension $d$. Since the cardinality of any sparse feature's set of indices, $S_{X_i}$, may be arbitrarily large, we apply a hash function $h_{m_i}: S_{X_i} \rightarrow \{0, 1, \ldots m-1\}$ over sparse feature indices. Consequently, the total parameters learned for all embedding for such a sparse feature is in the order of $d \times m_i$. It's worth noting that while we use a fixed embedding size $d$ for all sparse features, 
the hash size $m_i$ can vary for each sparse feature. 

\begin{figure*}[t] 
\centering
\includegraphics[width=0.32\textwidth]{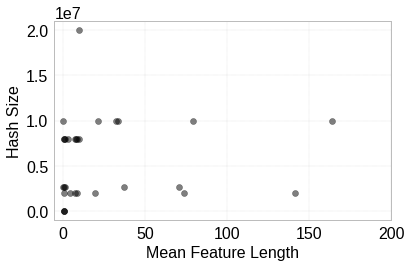} 
\includegraphics[width=0.32\textwidth]{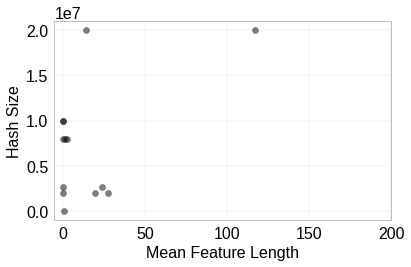} 
\includegraphics[width=0.32\textwidth]{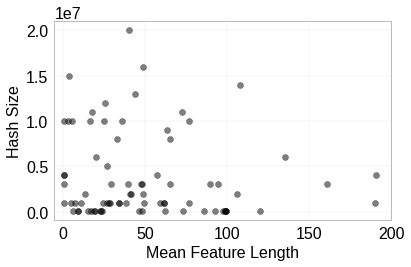} 
\caption{Hash size vs mean feature length of embedding tables in M1,2,3\textsubscript{prod} (from left to right).}
\vspace{-0.1cm}
\label{feature_size_dist}
\end{figure*}

\begin{figure*}[t] 
\centering
\includegraphics[width=0.32\textwidth]{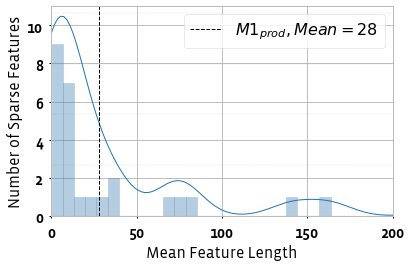} 
\includegraphics[width=0.32\textwidth]{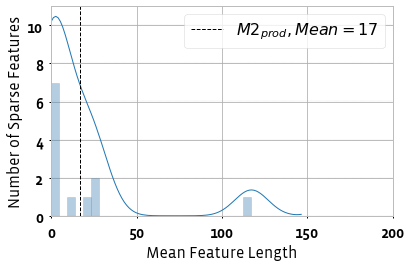} 
\includegraphics[width=0.32\textwidth]{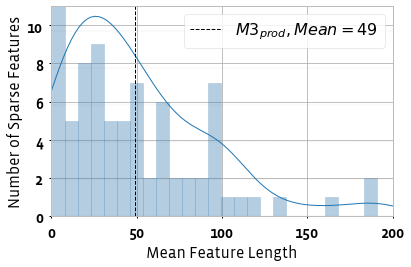} 
\caption{Mean sparse feature length distributions of M1,2,3\textsubscript{prod} (from left to right) with lines showing the kernel density estimates (KDE) that represents continuous probability density curve.}
\vspace{-0.2cm}
\label{feature_length_dist}
\end{figure*}

\subsubsection{\textbf{Embedding Tables}} Categorical input features, 
i.e. sparse features, determine how many embedding tables there will 
be. Given the large size and sparsity
of these tables, sparse features can be configured to share embedding 
tables to reduce the overall size of the model.
Since this requires a shared hash sizing for sharing the same 
embedding lookup table, this is generally useful for semantically 
similar sparse features. We expect similar sparse features
should have similar dense
representations (e.g. \textit{current item ID}, and \textit{last n 
interacted item IDs}).

\textbf{Embedding Table Hash Size.} Various methods can be used to limit the size of large embedding tables~\cite{ginart2019mixed},
and hashing is a common method for this.
Hash size is a customizable parameter per sparse feature.
The hash size $m_i$ varies depending on the semantics of $X_i$. For 
example, we would set a small $m_i$ if $X_i$ is indexed on towns 
within a given county, but set a much larger $m_i$ if $X_i$ is indexed on all known plant species.
Figure~\ref{feature_size_dist} shows the hash sizes
of the three different production models at Facebook.
Hash sizes for these models range from 30 being smallest, to 20 million the largest. Average hash
size for these models are 5.7 million, 7.3 million and 3.7 million
respectively.

Due to collisions hashing algorithms create,
lower hash sizes might cause accuracy degradation, while
providing the benefit of reducing the embedding table sizes.
In an earlier work, no hashing mechanism was used in order to avoid 
any model quality regression that was caused by compressing the 
embedding tables~\cite{zhao2020distributed}.
This results in tables in the order of terabytes.
Therefore, hash sizes of the features effect the
embedding table placement strategy, which we elaborate later.




\textbf{Embedding Table Access and Number of Look-ups per Table.} For a given input example in the training set, every sparse feature is a one-hot or multi-hot vector that has some arbitrary $n$ number of \textit{activated} indices with non-zero values. For each activated index, we perform an embedding table lookup, resulting in
fetching of $n$ embedding vectors. The $n$ embedding vectors are aggregated or pooled for the example's embedding representation for the specific sparse feature.
While the average cost of the embedding operations increase 
linearly with the average number of activated indices for each sparse 
feature, we may bound this operation with an optional 
\textit{truncation size} to limit the outliers.

We characterize feature length distributions (mean lookup 
operations per feature) of three production models
in Figure~\ref{feature_length_dist}.
Furthermore, we illustrate the correlation between
feature length distributions and embedding table hash sizes in Figure~\ref{feature_size_dist}.
Feature length distribution resembles a power-law distribution 
for each of the three models, i.e. there exists a small number of 
tables that are accessed much more frequently than others. 
Differences in access ratios might create imbalances among servers
if not carefully partitioned.
Furthermore, the access frequency does not always correlate with 
the embedding table size -- some of the most accessed tables 
are relatively small. The characterization results on the embedding
tables open up new optimization opportunities as well, such
as caching~\cite{zhao2020distributed} and compression for these 
large embedding tables using quantization~\cite{ginart2019mixed}.

\subsubsection{\textbf{Feature Interaction}} Outputs of dense and
sparse feature embedding operations are combined using functions,
such as, concatenation or pairwise dot product. In the case of concatenation, pooled embeddings of each sparse feature is
concatenated to the output of the dense MLP layer.
Alternatively, a pairwise dot product combiner helps in capturing
feature interactions between dense and sparse features, and also
among  sparse features.
Here, we can project the dense MLP output layer to a 
set of embeddings of dimension $d$ and compute the dot product between pairs of dense projections and sparse embeddings to enable sparse-dense interactions.
We can also compute the dot product between pairs of sparse embeddings to capture sparse-sparse interactions. These resulting dot products may be concatenated with the original dense MLP output layer and used as the input for the top MLP stack of the recommendation model.

\begin{table*}[ht]
\centering
\begin{tabular}{l|c|c|cl}
                            & \multicolumn{1}{l|}{\textbf{CPU System}} & \textbf{Big Basin GPU System} & \textbf{ Prototype Zion GPU System} &  \\ \cline{1-4}
\textbf{Accelerators}       & -                                        & 8 NVIDIA V100           & 8  NVIDIA V100          &  \\ \cline{1-4}
\textbf{Accelerator Memory} & -                                        & 16/32 GB                & 32 GB                   &  \\ \cline{1-4}
\textbf{System Memory}      & 256 GB                                   & 256 GB                  & $\sim$2 TB              &  \\ \cline{1-4}
\textbf{CPU}                & 2 sockets                       & 2 socket, 20 cores      & 8 socket CPU   &  \\ \cline{1-4}
\textbf{Interconnect}       & 25 Gbps Ethernet	                       & 100 Gbps Ethernet       & 4X Infiniband 100 Gbps  & 
\end{tabular}
\caption{Hardware platform details.}
\vspace{-0.2cm}
\label{tab:hardware}
\end{table*}

\subsubsection{\textbf{MLP Dimensions}} There are two major MLP stacks in deep learning recommendation models: the dense feature MLP stack at the bottom of the recommendation model and the top MLP stack at the end. The dense feature MLP stack is applied on top of the dense input in order to reduce the thousands of input features into a much denser vector (e.g. 64 or 128 elements). The top MLP stack is applied after the feature interaction. Both \textit{depth} and \textit{width} of the MLP
stacks can be tuned across different model training runs.

\subsubsection{\textbf{Batch Size}} Batch size is a critical hyper-parameter
that affects training performance and model quality~\cite{goyal2017accurate}.
We usually scale the batch size as a function of the training system's capacity for parallelization. Typically, distributed training on CPUs uses a much smaller batch size per CPU, i.e. mini-batch size, relative to GPUs.
On the other hand, GPUs require higher mini-batch sizes to fully utilize the GPU compute capacity. 
Training throughput increases roughly linearly with the increasing batch size, up to a limit. 
We discuss the batch-size throughput scaling further in Section~\ref{efficiency}. 
    

\subsubsection{\textbf{Gradient Synchronization Method}} For distributed training, model parameters need to be synchronized across trainers and/or parameter servers. Gradient synchronization method is an important factor affecting both the model quality as well as scaling the performance.
There are two main approaches for synchronizing gradients: synchronous ~\cite{goyal2017accurate,akiba2017extremely,lian2017decentralized,jiang2017collaborative} and asynchronous~\cite{dean:nips2012,lian2017asynchronous,jin2016scale}.
Most commonly-used asynchronous algorithms
in deep learning recommendation model training
include Elastic-Averaging SGD (EASGD)~\cite{zhang2015deep}, Hogwild!~\cite{NIPS2011_4390} and Facebook's ShadowSync algorithm~\cite{zheng2020shadowsync}.

\section{Hardware and System Configurations} 
\label{hardware}

After model input and architecture configurations are made, next step is to decide on hardware and system configurations. There is often a de-facto hardware type to go for each workflow. However, as model configurations change, most efficient hardware choice and system configurations could also change over time. Here we give details on some of the hardware platforms in our heterogeneous datacenters and explain the major system configurations.

\subsection{Hardware Platforms}
Training workflows at Facebook leverage a large fleet of CPU and GPU 
platforms to serve necessary training frequencies at required
service latency. 
Three of these platforms, which are used in this paper, are summarized in Table~\ref{tab:hardware}.
Designs for each of these platforms have been publicly released through the Open Compute Project.

\textbf{Dual-Socket CPU} platforms house two Intel Skylake CPUs with 256 GB of DRAM~\cite{hazelwood:hpca2018}.

\textbf{Big Basin GPU} platform  has  two  Intel  CPUs  (various  generations)  and  eight  NVIDIA  GPUs~\cite{big_basin}. The Tesla V100  GPU accelerators  are  connected  using  NVIDIA  NVLink  to  form  an eight-GPU  hybrid  cube mesh.
The  V100  platform  enables 15.7 teraflops of single precision floating-point arithmetic per GPU. It also has high-bandwidth  memory  (HBM2)  providing 900GB/s  bandwidth. The  GPUs  are equipped with either 16 or 32 GB of memory where as the CPUs are with 256 GB DRAM and are connected via 100 Gbps Ethernet. 

\textbf{Zion GPU} platform has 8 CPU sockets 
interconnecting with the 8 GPU accelerators to provide the high compute and memory capacity~\cite{naumov:arxiv2020,zion}. Compared with the Big Basin GPU platform, accelerators are the same but
the system memory capacity, bandwidth and CPU compute capacity of Zion
is much larger, with $\sim$2 TB system memory and $\sim$1 TB/s memory bandwidth.

\begin{figure}[t]
\centering
\includegraphics[width=\columnwidth]{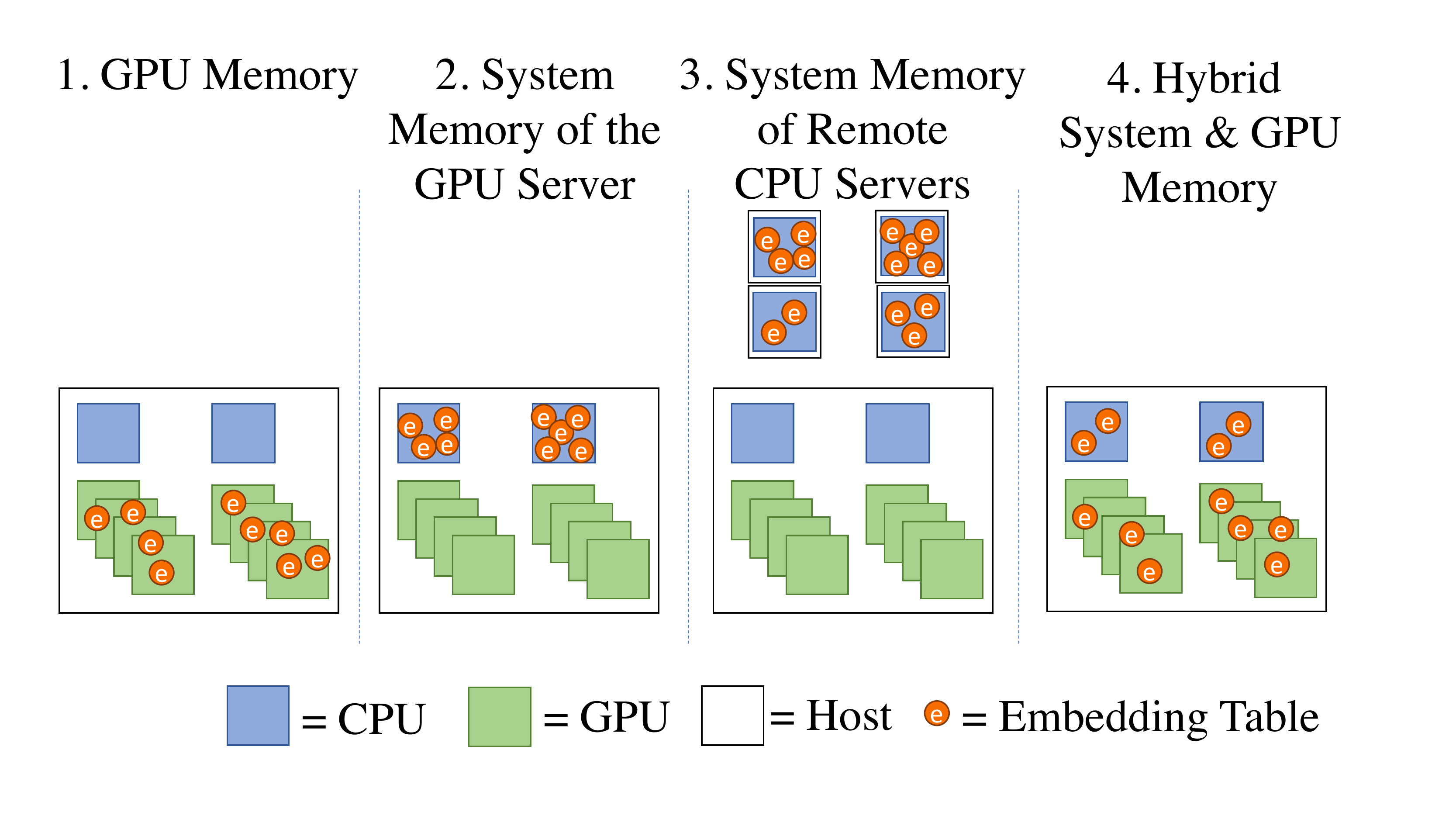}
\caption{Embedding table placement options.}
\vspace{-0.3cm}
\label{embed_placement}
\end{figure}

\subsection{System Configurations}
Each hardware requires different system configurations and these configurations affect the underlying software framework. Two of the major configurations we discuss in this section
are embedding table placement and selection of number of servers. 
Other system configuration options include tuning number of worker threads, data loading threads.

\subsubsection{\textbf{Embedding Table Placement}}
For training on CPU-only servers, embedding tables can be simply placed on system
memory. For accelerated systems, the optimal embedding placement strategy might differ given the hardware properties.
We discuss four strategies for storing embedding tables for GPU platforms: GPU memory, system memory of the GPU server,
system memory of remote CPU servers, hybrid system and GPU memory.
These are visualized in Figure~\ref{embed_placement}.
Other alternative placement options studied in the literature include storing the embedding tables in non-volatile memory~\cite{eisenman2018bandana} and SSDs~\cite{zhao2020distributed}.

\textbf{On the GPU memory.} Embedding tables are distributed among GPUs, different partitioning strategies can be used such as table-wise or row-wise partitioning. For GPU servers that have high-bandwidth inter-GPU communication, like Big Basin, storing embedding tables on the GPUs enables offloading all model operations to be done on the GPU, minimizing the CPU usage and CPU-GPU copy operations. Big Basin design includes eight NVIDIA GPUs connected using high-speed interconnect NVLink and contains 128/256 GB total available GPU memory per node to store the embedding tables. For models that do not fit on a single server, inter-node
communication (i.e. interconnect) becomes an important factor affecting performance.

\textbf{System memory of the GPU server.} Storing the embedding tables on the system memory of the CPUs of the GPU server would be a good option for servers with large system memory and high memory bandwidth. For example, a training platform like Zion contains, 8 socket CPUs, 2 TB memory with ~1 TB/s memory bandwidth as shown in Table~\ref{tab:hardware}. In contrast, a platform like Big Basin contains 2 socket CPUs for 8 accelerators with 256 GB system memory. For this setup, CPUs are likely to become a bottleneck. This option also makes it more difficult to use the GPUs effectively for all operators since the data is not located on the GPU memory.

\textbf{System memory of remote CPU servers.} On a platform like Big Basin, storing the embedding tables in the remote CPU system memory enables scaling out the number of CPUs used for embedding table storage and might remove the CPU bottleneck mentioned in option 2. However, this increases the latency of the sparse lookup operations since they have to go through the network. This setup also creates additional work for the CPUs on the GPU server to do the remote send/receive operations and CPU to GPU data movement.

\textbf{Mixed system and GPU memory.} This is a hybrid placement strategy where some of the embedding tables are stored on the GPUs and some are stored on the system memory. This can be an alternative strategy when the embedding tables do not fit on the GPU, placing as much as tables as it can fit could reduce the pressure on the CPU.

\begin{figure}[t]
\centering
\includegraphics[width=0.9\columnwidth]{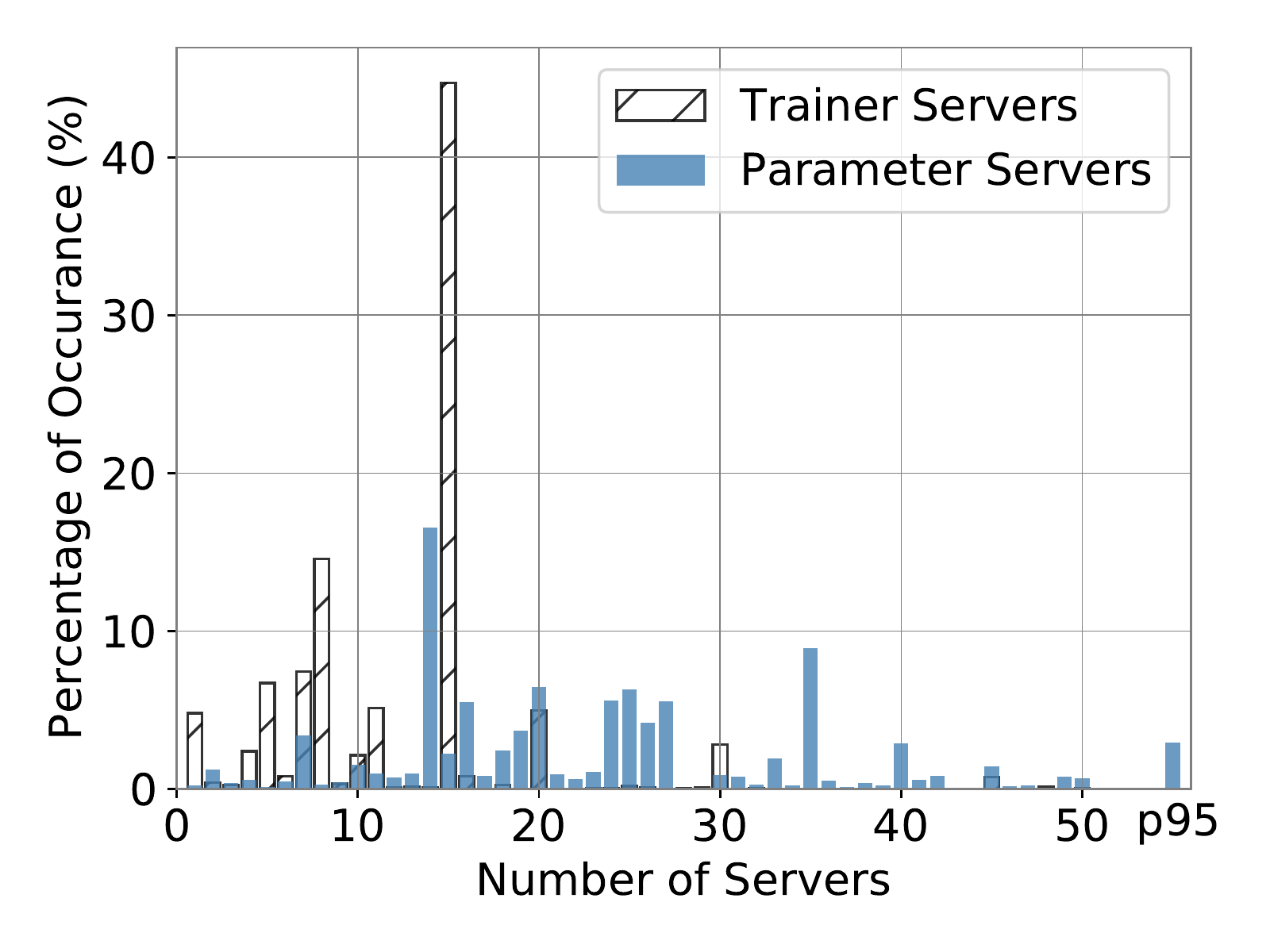}
\vspace{-0.2cm}
\caption{Histogram of number of trainer and parameter servers used for training ranking models over a month period.}
\vspace{-0.25cm}
\label{server_dist}
\end{figure}

\begin{figure*}[t] 
\centering
\includegraphics[width=0.32\textwidth]{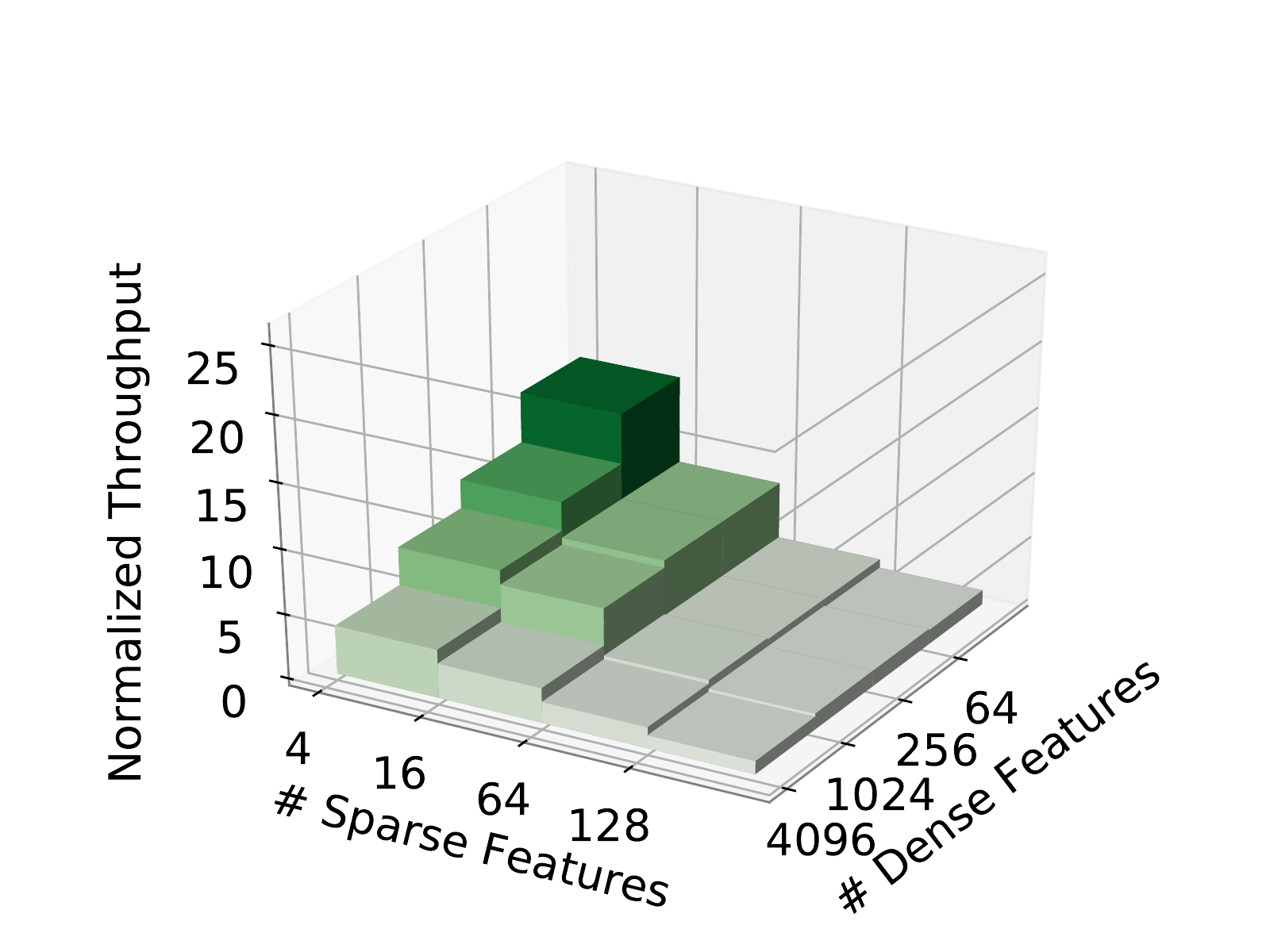}
\includegraphics[width=0.32\textwidth]{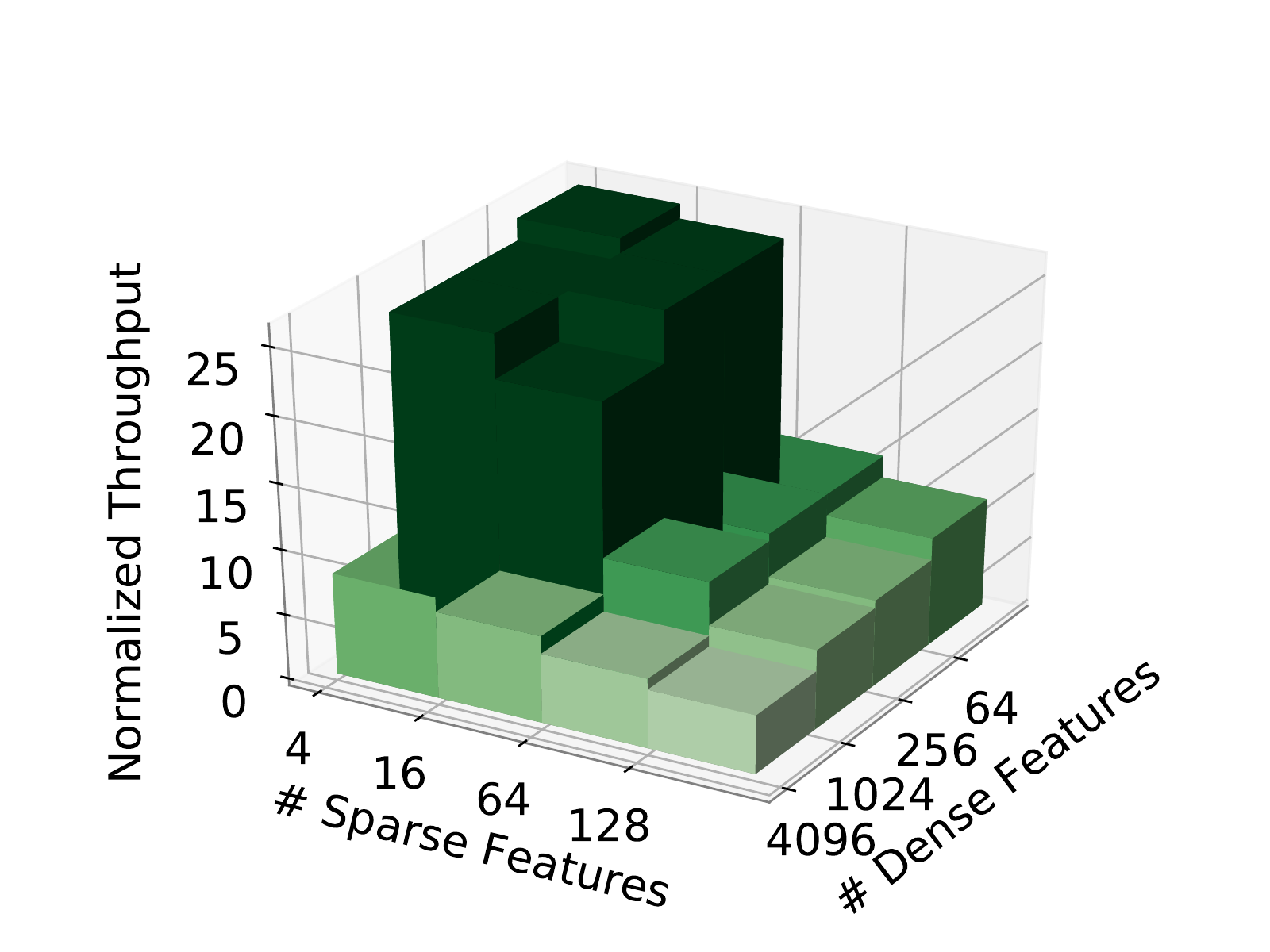}
\includegraphics[width=0.33\textwidth]{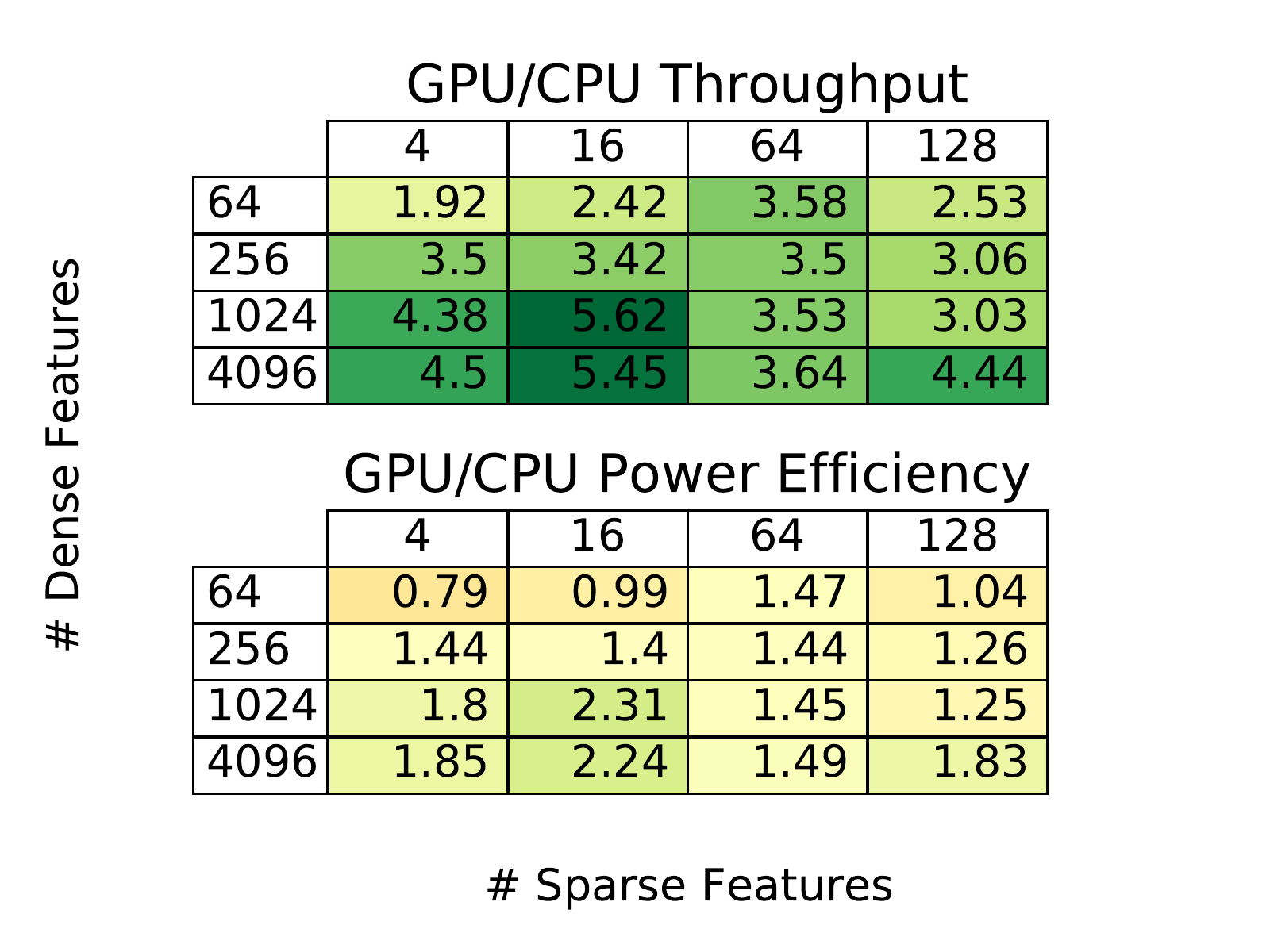} 
\caption{Varying the number of sparse/dense features on CPU (left), GPU (middle) and system efficiency comparison (right). Fixed MLP dimensions of $512^3$, hash size of $100000$ and batch sizes of $200$ (for CPU), $1600$ (for GPU) are used.}
\vspace{-0.1cm}
\label{test_suite_features}
\end{figure*}

\subsubsection{\textbf{Number of Servers}} Numbers of three types of
servers need to be configured for each workflow: number of trainer 
servers, number of parameter servers (if used)
and number of reader servers.
Selection of number of servers is made based on the
throughput requirement for training and the memory capacity requirement of the
embedding tables.
Further the numbers can vary significantly from run-to-run.

\textbf{Trainer Servers.} Trainer servers dictate the data parallelism with which we train. Generally, we see an approximately linear increase in training speedup when we increase the number of trainer servers, up to
a certain degree. However extra trainers can introduce model quality loss during model parameter synchronization in training.
Figure~\ref{server_dist} shows the distribution of number of trainers used
in the datacenter over a month long period for the CPU training
workflows.
As the training throughput requirement does not change
very often, this leads to over 40\% of the workflows using
same number of trainers.

\textbf{Parameter Servers.} Parameter servers hold the model parameters in memory. These are partitioned between dense and sparse servers, wherein MLP layers are stored on dense servers and
embedding tables for sparse feature are stored on sparse servers.
Distribution of number of parameter servers used for the training workflows that run on CPUs is shown in Figure~\ref{server_dist}.
In contrast to number of trainers, number of parameter servers
vary greatly.
As the ML engineers experiment with different features,
memory capacity requirement changes frequently
which results in a wide-range of number of parameter servers.

\textbf{Reader Servers.} Readers access model training data 
in parallel from remote storage Hive, Facebook’s exabyte-scale data warehouse~\cite{thusoo2010data}. Reader servers are decoupled from
trainers to be scaled-up independently and not to stall training.
We typically scale up reader servers such that data reading is not a bottleneck. Consequently, for more performant training hardware, we may utilize more readers.

\section{Efficiency of model setup configurations}
\label{efficiency}
\begin{figure*}[t] 
\centering
\includegraphics[width=0.45\textwidth]{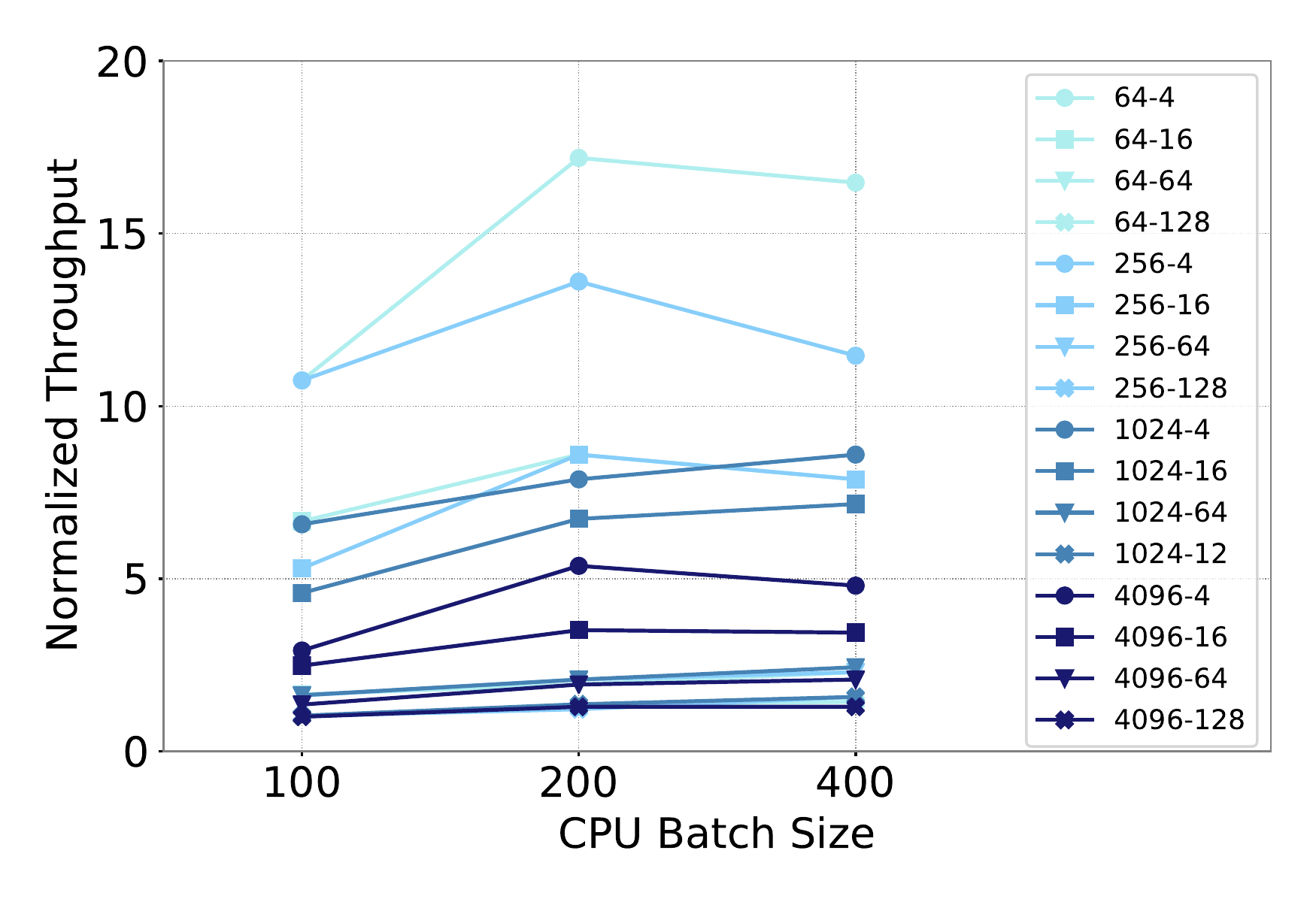}
\includegraphics[width=0.45\textwidth]{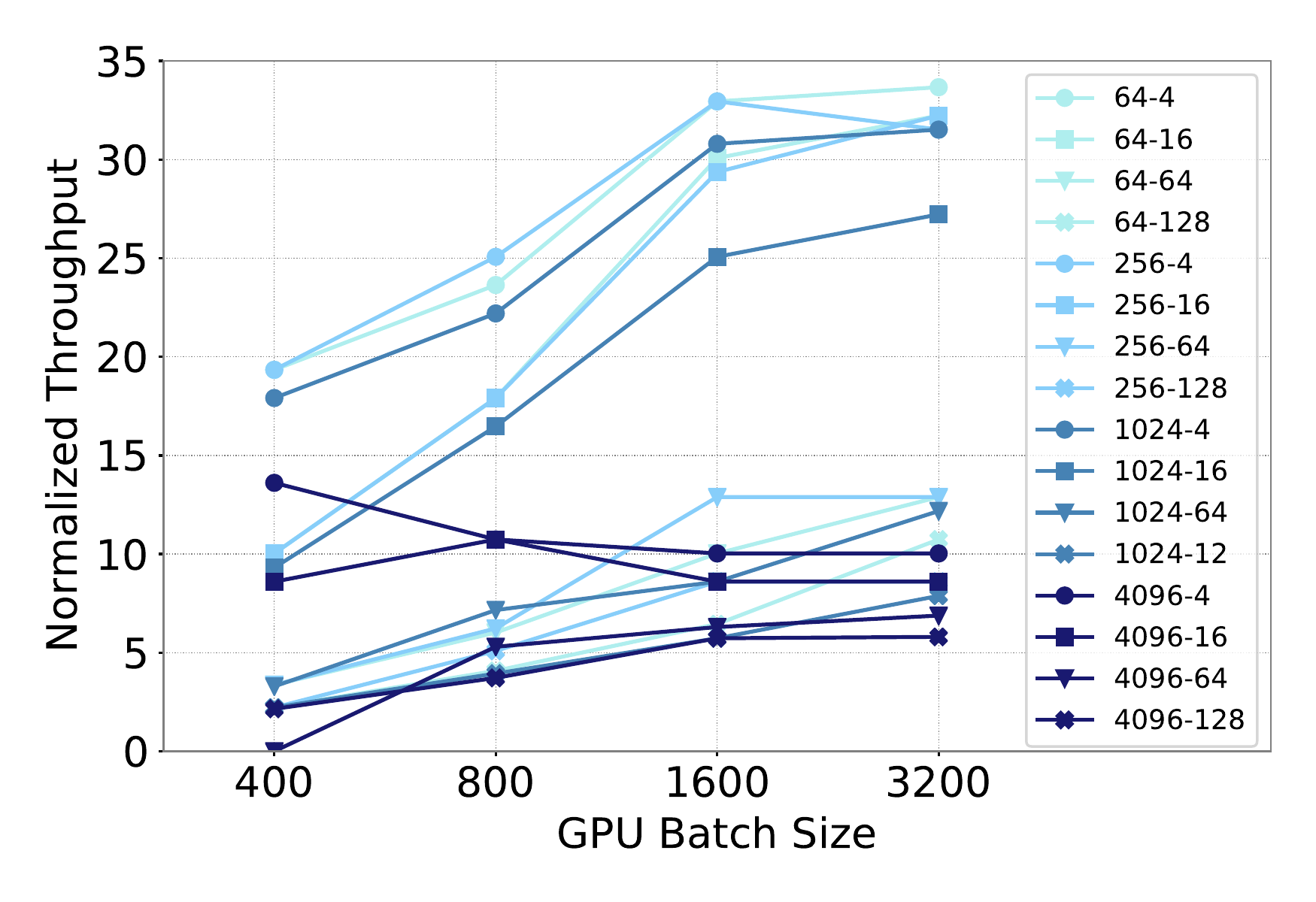}
\vspace{-0.1cm}
\caption{Batch size scaling on CPU (left) and GPU (right). Fixed MLP dimensions of $512^3$ and hash size of $100000$ are used.}
\label{test_suite_batch_size}
\end{figure*}

The parameter space of all the possible model and system architecture
combinations is exponential and therefore challenging to study.
We study a subset of the most important model configurations in a
parameterized manner to understand their effects on throughput and to
compare their behavior on CPU and GPU systems.
With this goal, we created a test suite where we can customize
major model configurations in a systematic way.

\textbf{Design Space Exploration:}
To explore the design space of training model configurations, we created
a model containing basic components of recommendation models as shown
earlier in Figure~\ref{training_configs}.
We configure our test suite setups to train over a
variety of numbers of dense features between 64 and 4096. Like dense features, we configure sparse features in our model training setups by testing counts of sparse
features ranging between 4 and 128, which is representative of the
typical bounds of the total number
of sparse features used in recommendation models at Facebook.
We fix a constant hash size for all sparse features in our model to remove potential noise added by training over sparse features with varying indexing. We truncate number of look-ups per table to 32, to limit outliers in our study.

\subsection{\textbf{Number of Sparse and Dense Features:} As the number of dense and sparse features increase, training throughput reduces because of the increasing memory overhead from embedding operations. Big Basin provides higher training throughput despite, in a few cases, with lower performance-per-watt energy efficiency.}
The number of features used in production models may vary heavily. This degree of variability can lead to very large discrepancies in model complexity, size, and time to train over some fixed set of examples.  Increasing the number of dense features leads to increases in the computational complexity of the first layer within the model's dense architecture. On the other hand, increasing sparse features leads to additional embedding lookup operations, pooling and interactions among large embeddings, all of which are computationally expensive and can lead to significant model size increases due to additional embedding tables. Not only does this increase the computational cost, but can also strain system memory.

Figure~\ref{test_suite_features} shows the throughput under CPU and GPU with varying dense and sparse feature combinations. In the CPU setup, we use a single trainer, dense and sparse parameter server. In the GPU setup, we use a single Big Basin GPU server where sparse and dense features are placed in GPU Memory. Power capacity requirement of a Big Basin server is $7.3$ times higher
than the dual-socket CPU server. We see that the throughput of the GPU setup is
higher than the CPU setup in all configurations. However, in terms of
power efficiency GPU does not always provide the best results.
GPU efficiency is the highest for models with more dense features. Note that
in this setup, we use fixed batch sizes of $200$ (for CPU), $1600$ (for GPU), a hash size of $100000$ (which is smaller compared to the average hash size of the production models) and MLP dimensions of $512^3$. Later in this section, we show the effect of scaling each of these parameters on throughput.

\subsection{\textbf{Batch size:} There exists a different throughput-optimal batch sizes for CPU and GPU training. The optimal batch size varies based on the sparse and dense configurations of the model.}

The batch size of the model dictates the number of examples processed
through a forward/backward pass during model training. Increasing the 
batch size can lead to a better training stability due to more accurate
estimates of the loss gradient. However, it can also lead to worse
model generalization. Increasing batch size can have varying effects on training throughput too. Higher batch sizes can be detrimental to the training speed over CPU hardware. In contrast, GPU hardware with significantly greater parallel compute capacity can better leverage greater batch sizes. We often see an approximately linear increase in training throughput, until a certain point.

Figure~\ref{test_suite_batch_size} shows the throughput
under CPU and GPU as we scale the batch size. As the batch size increase, 
communication per iteration increases since the number of embedding lookup operations per iteration also increase.
For GPU training, large batch size reduces the overhead from CUDA API calls such as kernel launches, as well as allowing better use of GPU compute capacity.
When communication overhead becomes larger than the GPU compute benefit, throughput starts to saturate. Even if increasing the batch size beyond
the first saturation point does not affect the throughput significantly,
it impacts the model quality. As we discuss later, given the importance of model quality in recommendation models, batch size needs to be
carefully selected for each model configuration.

\begin{figure*}[h]
\centering
\includegraphics[width=0.45\textwidth]{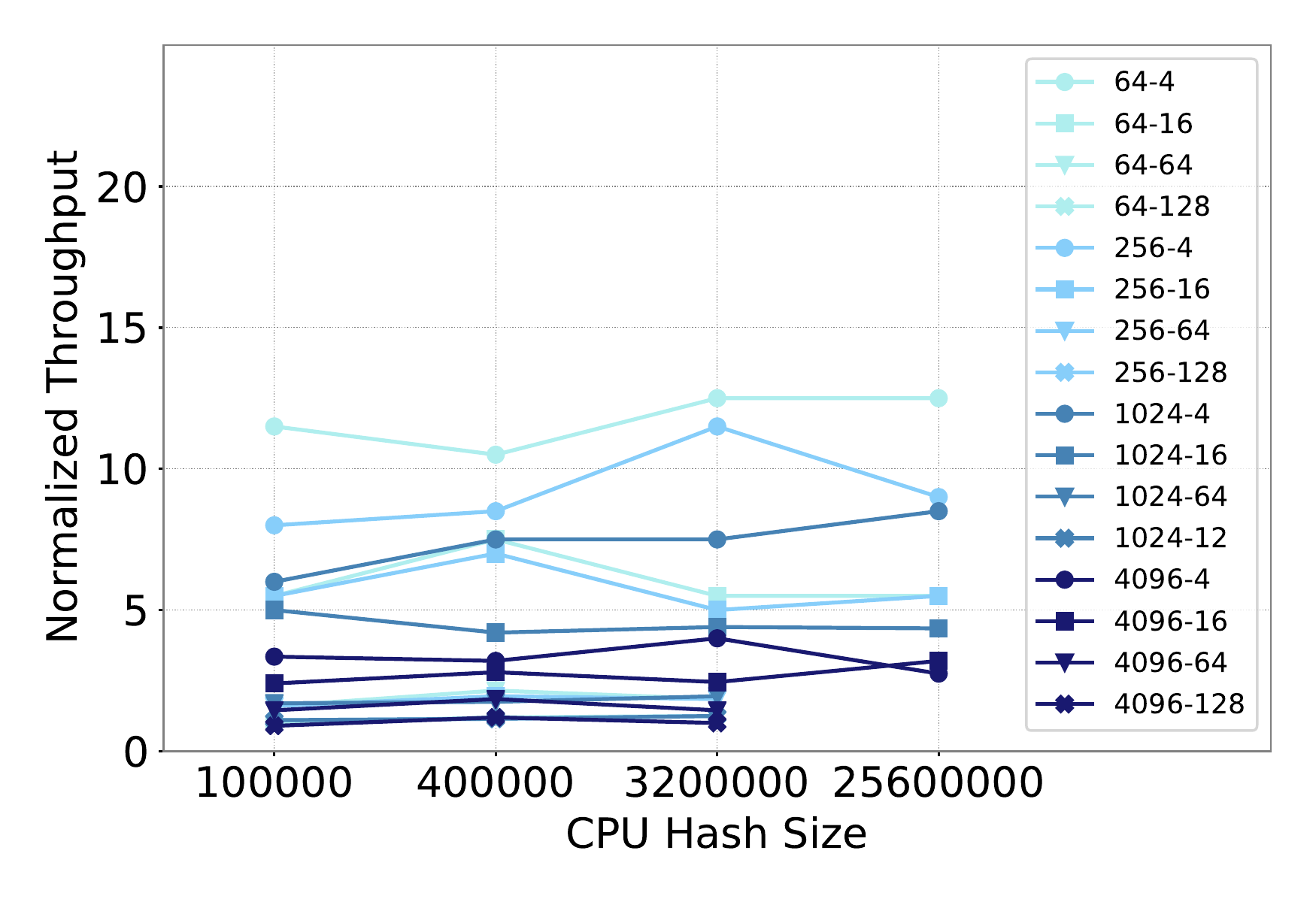}
\includegraphics[width=0.45\textwidth]{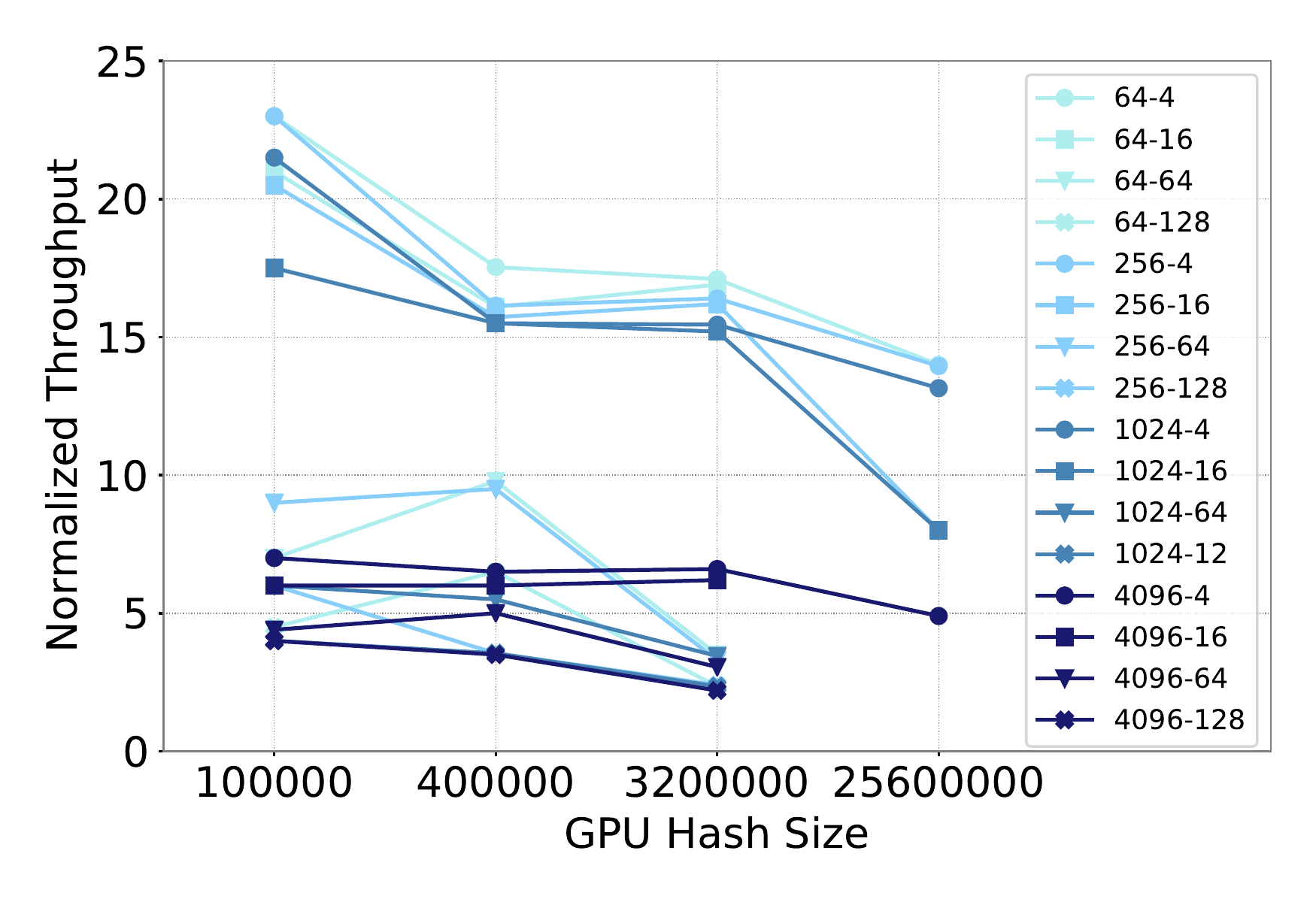}
\vspace{-0.1cm}
\caption{Hash size scaling on CPU (left) and GPU (right).}
\vspace{-0.3cm}
\label{test_suite_hash_size}
\end{figure*}

\subsection{\textbf{Embedding Table Hash Size:} As the effective size of embedding tables grows with an increasing hash size, the training throughput of Big Basin reduces significantly due to the increasing communication overhead between the GPUs.}
Each sparse feature's hash size denotes the number of 
entries of that feature's embedding lookup table, wherein each entry is a $d$-dimensional embedding. Increasing this parameter yields a linear increase in the embedding table size and the total model size, but does not significantly affect embedding lookup time or overall model throughput on CPU. For GPU training, due to the limited HBM2 memory of the GPU compared with the system memory of the CPU, more GPUs need to be used in order to store the embeddings.

Figure~\ref{test_suite_hash_size} shows the throughput
with varying hash size on CPU and GPU. We use a single CPU
parameter server with 256 GB memory and a single Big Basin
server with 256 GB GPU memory for a fair comparison. During CPU training, all tables are stored in the same single server
as the hash size increase. During GPU training, as the hash size increase more GPUs within the single server need to be used to fit the increasing
size of the embedding tables and this increases the
communication cost between GPUs. Therefore, throughput drops 
significantly as we scale hash size.

\subsection{\textbf{MLP Dimensions:} Increasing the width of the MLP layers and the number of sequential layers reduces CPU training throughput higher than the GPU throughput.}
\begin{figure}[htb!] 
\centering
\includegraphics[width=0.40\textwidth]{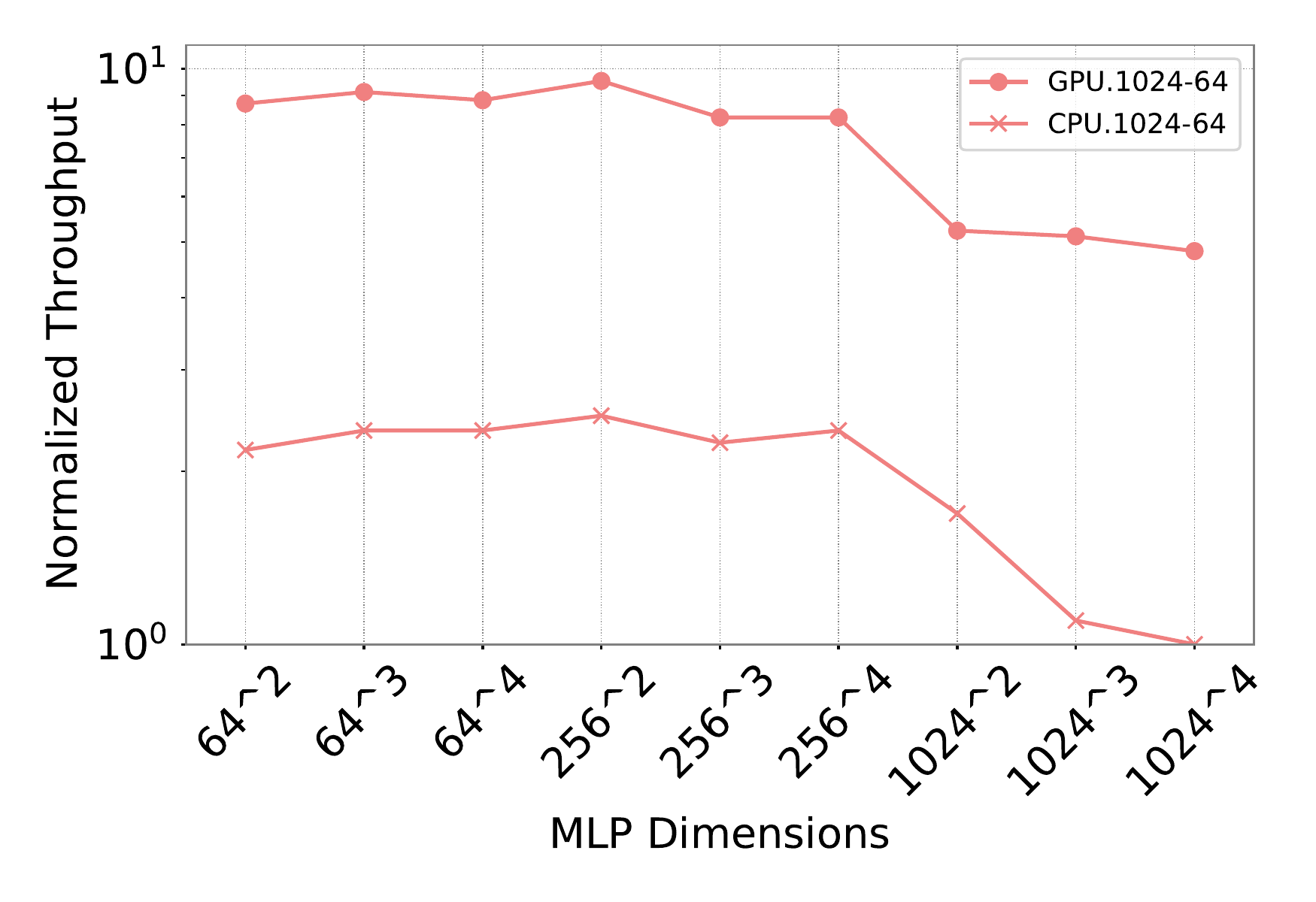}
\vspace{-0.1cm}
\caption{Throughput under varying MLP dimensions.}
\vspace{-0.4cm}
\label{test_suite_mlp}
\end{figure}

Figure~\ref{test_suite_mlp} shows the training throughput under varying
MLP dimensions on CPU and GPU.
We denote the MLP width and layers as $width^{num\_layers}$, e.g. $64^2$ means 2 layers of size 64.
For smaller MLP dimensions, the relatively high number of sparse features in the model resulted in disproportionately more expensive embedding lookup, pooling and interaction. Therefore, we do not see the throughput decrease significantly until the MLP dimension grows larger than ($256^3$) for both CPU and GPU training. When the MLP dimensions grow, the drop in normalized relative throughput is higher
for CPU training compared with GPU training due to the higher compute 
capacity of the GPU.

\section{Case Study with production-scale models}
\label{case_studies}

\begin{table}
\centering
\tabcolsep=0.11cm
\begin{tabular}{|l|c|c|c|} 
\cline{1-4}
                                 & \textbf{M1\textsubscript{prod}} & \textbf{M2\textsubscript{prod}} & \textbf{M3\textsubscript{prod}} \\ \cline{1-4}
\textbf{\# Sparse Features}      & 30            & 13            & 127           \\ \cline{1-4}
\textbf{\# Dense Features}       & 800           & 504           & 809           \\ \cline{1-4}
\textbf{Embedding Size {[}GB{]}} & tens          & tens          & hundreds      \\ \cline{1-4}
\textbf{Embedding Lookups}       & 28            & 17            & 49            \\ \cline{1-4}
\textbf{Bottom MLP Dimensions}  & 512 & 1024 & 512    \\ \cline{1-4}
\textbf{Top MLP Dimensions}  &  512-512-512 & 1024-1024-512 &      512-256-512 \\
 &   & &     -256-512 \\ \cline{1-4}
\end{tabular}
\caption{Descriptions of three production models.}
\label{tab:model_descriptions}
\end{table}

\begin{table*}[]
\tabcolsep=0.11cm
\centering
\begin{tabular}{|l|c|c|c|}
\cline{1-4}
& \textbf{M1\textsubscript{prod}}  & \textbf{M2\textsubscript{prod}}  & \textbf{M3\textsubscript{prod}}     \\ \cline{1-4}
\textbf{CPU Setup} & \begin{tabular}[c]{@{}c@{}}6 trainers\\ 8 parameter servers\end{tabular} & \begin{tabular}[c]{@{}c@{}}20 trainers\\ 16 parameter servers\end{tabular} & \begin{tabular}[c]{@{}c@{}}8 trainer\\ 8 parameter servers\end{tabular}   \\ \cline{1-4}
\textbf{GPU Setup} & 1 Big Basin GPU & 1 Big Basin GPU & 1 Big Basin GPU   \\ \cline{1-4}
\textbf{Embedding Placement} & GPU Memory & GPU Memory & Remote CPU Memory  \\ \cline{1-4}
\textbf{Sync Mode} & easgd, 1 hogwild  & easgd, 1 hogwild  & easgd, 4 hogwild  \\ \cline{1-4}
\textbf{Optimal Batch Size per GPU} & 1600 & 3200 & 800   \\ \cline{1-4}
\textbf{GPU/CPU Relative Throughput} & \cellcolor[HTML]{D4F8D4}2.25x & \cellcolor[HTML]{F5F5BC}0.85x & \cellcolor[HTML]{FFCCC9}0.67x  \\ \cline{1-4}
\textbf{GPU/CPU Power Efficiency} & \cellcolor[HTML]{D4F8D4}4.3x & \cellcolor[HTML]{D4F8D4}2.8x  & \cellcolor[HTML]{FFCCC9}0.43x   \\ \cline{1-4}
\end{tabular}
\caption{CPU-GPU optimal setup comparison.}
\vspace{-5mm}
\label{tab:case_study_results}
\end{table*}

We present case studies for three production 
recommendation models (M1\textsubscript{prod}, M2\textsubscript{prod}, and M3\textsubscript{prod})
to compare CPU and GPU efficiency in real-life scenarios.
These models are written in the Caffe2 framework, which is now a part of PyTorch~\cite{caffe2}, 
and use FP32 (single precision floating point) precision.
Table~\ref{tab:model_descriptions} summarizes the main 
properties of the three models evaluated here. 
The first two models (M1\textsubscript{prod} \& 2)
have lower number of sparse features than the third one. 
In terms of dense features, M2\textsubscript{prod} has lower number of 
dense features than the others.
Moreover, in respect of embedding sizes, M1\textsubscript{prod} \& 2 have embedding tables 
in the order of tens of GBs, and M3\textsubscript{prod} has in the order of hundreds.
Two important properties captured by the production models that
were not captured by the test suite are: the varying number of lookups per table and the diverse hash sizes
for the individual embedding tables as we presented earlier in Figure~\ref{feature_size_dist} and~\ref{feature_length_dist}.

\subsection{\textbf{CPU vs Big Basin GPU Performance:}
When
embedding tables fit on the GPU memory, as compared to CPUs, Big Basin
provided higher throughput and power efficiency.}

We compare the relative throughput and power efficiency between the CPU and Big Basin GPU training systems.
Table~\ref{tab:case_study_results} shows the original production 
CPU setups and our prototype on Big Basin. 
When we port these models into GPU, the first step
is finding the optimal batch size as we discussed in the previous section. We found the throughput started to saturate after batch size
of 1600, 3200 and 800 for M1,2,3\textsubscript{prod}
respectively. We evaluated M1\textsubscript{prod} \& M2\textsubscript{prod} on a single Big Basin GPU and stored the embedding tables on GPU memory.

M1\textsubscript{prod} achieved 2.25 times higher throughput on GPU
compared to its production CPU setup and is 4.3 times
more power efficient. M2\textsubscript{prod} achieved close performance
to its production CPU setup (i.e. 0.85 times) and achieved
2.8 times higher power efficiency on Big Basin GPUs.
As we showed in the previous section, when embedding tables are placed
on GPU memory, throughput and efficiency exceeds the CPU setup in most cases.
On the other hand, when embedding tables do not fit on the limited
GPU memory, it is harder to achieve the same level of efficiency.
For M3\textsubscript{prod}, when embedding tables do not fit on the high-bandwidth
memory of the Big Basin GPUs, we use remote CPU servers to serve as 
parameter servers and a single Big Basin GPU server to serve as a trainer.
We scaled-up the number of parameter servers and yet the throughput
on Big Basin only reached 0.67x of the throughput of the
production setup on 8 CPU trainers. We found that CPU resources
on the Big Basin server and data copies from the parameter servers 
became the performance bottleneck.

When the embedding tables do not fit on a single GPU server,
another option is to distribute the tables on the GPU memory of multiple Big-Basins. To be performance efficient, this mode requires
fast inter-node GPU-GPU communication for embedding look-ups.
Due to the lack of this capability, we were not able to test
this model setup on multiple Big Basins.

\subsection{\textbf{Big Basin vs Zion GPU Performance:} Zion's large memory capacity, bandwidth and CPU compute
resources provide several orders of magnitude efficiency improvement
over Big Basin, when the embedding tables do not fit on GPU memory.}

\begin{figure}[t] 
\centering
\includegraphics[width=0.37\textwidth]{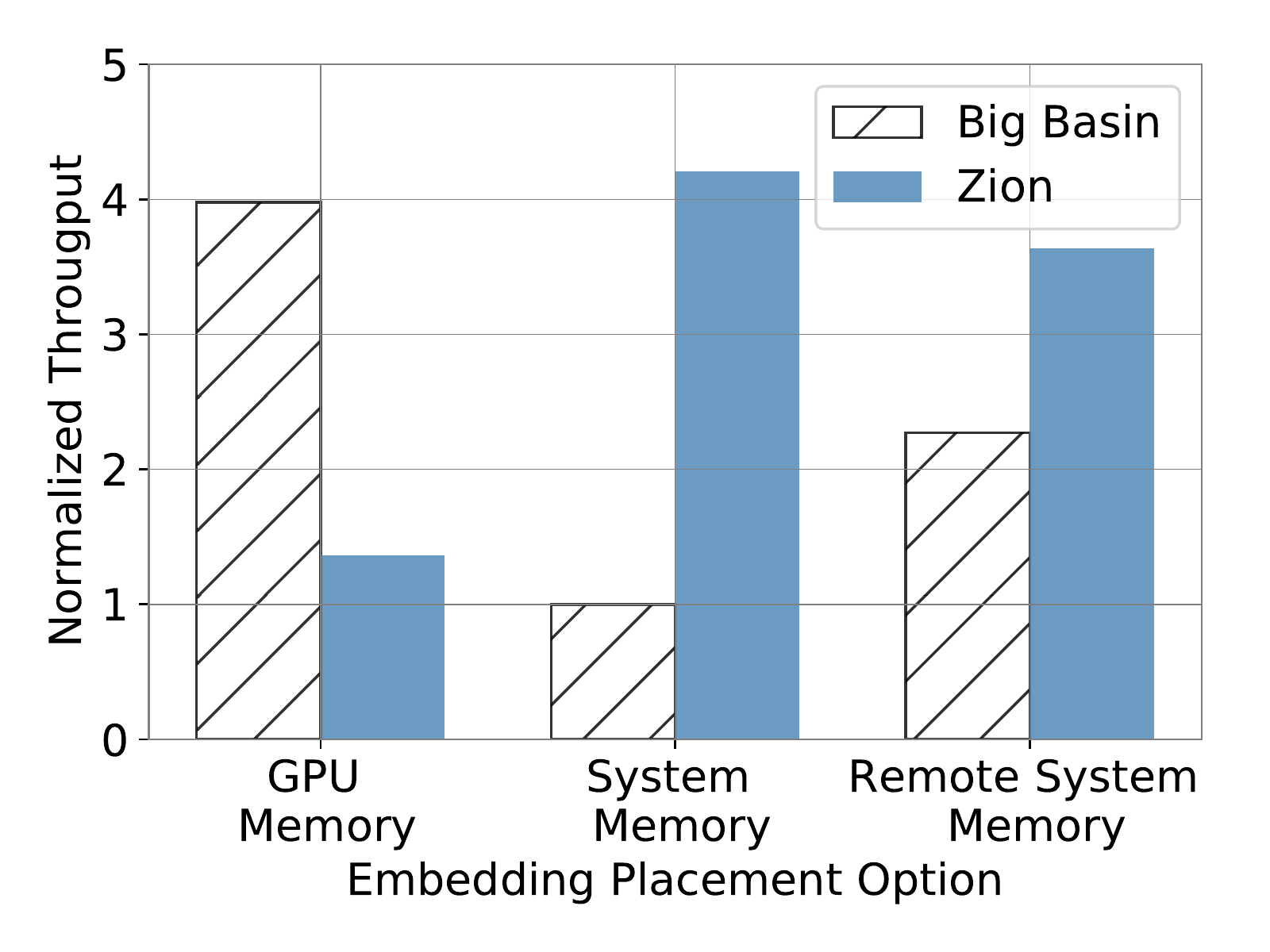}
\vspace{-0.1cm}
\caption{Embedding placements on Big Basin vs. Zion for M2\textsubscript{prod}.}
\vspace{-0.4cm}
\label{placement_zion_bbv}
\end{figure}

As model size and complexity increase,
improving the training throughput efficiently becomes a challenge.
Zion is Facebook's large-memory training platform trying to address this challenge.
We compare the training throughput with different placement options on Big Basin and Zion using one of the models (M2\textsubscript{prod}) on Figure~\ref{placement_zion_bbv}.
With GPU memory placement, Big Basin showed the best performance.
Zion's performance was much lower because there was no GPU-GPU direct communication in our \textit{prototype Zion server}, hence all communication across GPUs went through CPUs. This shows the importance of the GPU-to-GPU interconnect when placing the embedding tables on GPUs.
With system memory placement, Zion performed the best as expected due to its high CPU memory bandwidth.
For Big Basin, throughput was four times lower than the first placement option due to slower memory bandwidth and low CPU resources.
With remote system memory placement, performance could not exceed other approaches on both Big Basin and Zion despite scaling up the number of remote parameter servers to store
the embedding tables.
We found that lookup latency and the CPU resources on the GPU server becomes a bottleneck.
The performance of Zion was only slightly better than Big Basin due to its
higher memory bandwidth and more CPU resources.

Zion's main advantage over Big Basin comes into play when embedding
tables do not fit on the GPU memory of a single GPU-server.
Its $\sim$2 TB system memory and
$\sim$1 TB/s memory bandwidth provide large capacity to store the
embedding tables and fast look-up operations. Our analytical model
showed that for RM-3, training on Zion is several orders of magnitude
more efficient than using multiple Big Basins with embedding tables placed
on the GPU memory. The challenge remaining for Zion is the case where 
model sizes grow into multiple terabytes which requires scaling out on 
multiple Zion servers.

\subsection{\textbf{Accuracy:} For applications requiring very well-calibrated predictions, model accuracy loss in the orders of $\sim$0.1\% may not be tolerable for recommendation models to achieve a higher training throughput. 
In order to test and improve the model prediction quality, high volumes of data are used.
This also increases the length of the process of hyper-parameter tuning,
which is an important part of training given the accuracy requirement.}

\begin{figure}[t] 
\centering
\includegraphics[width=0.40\textwidth]{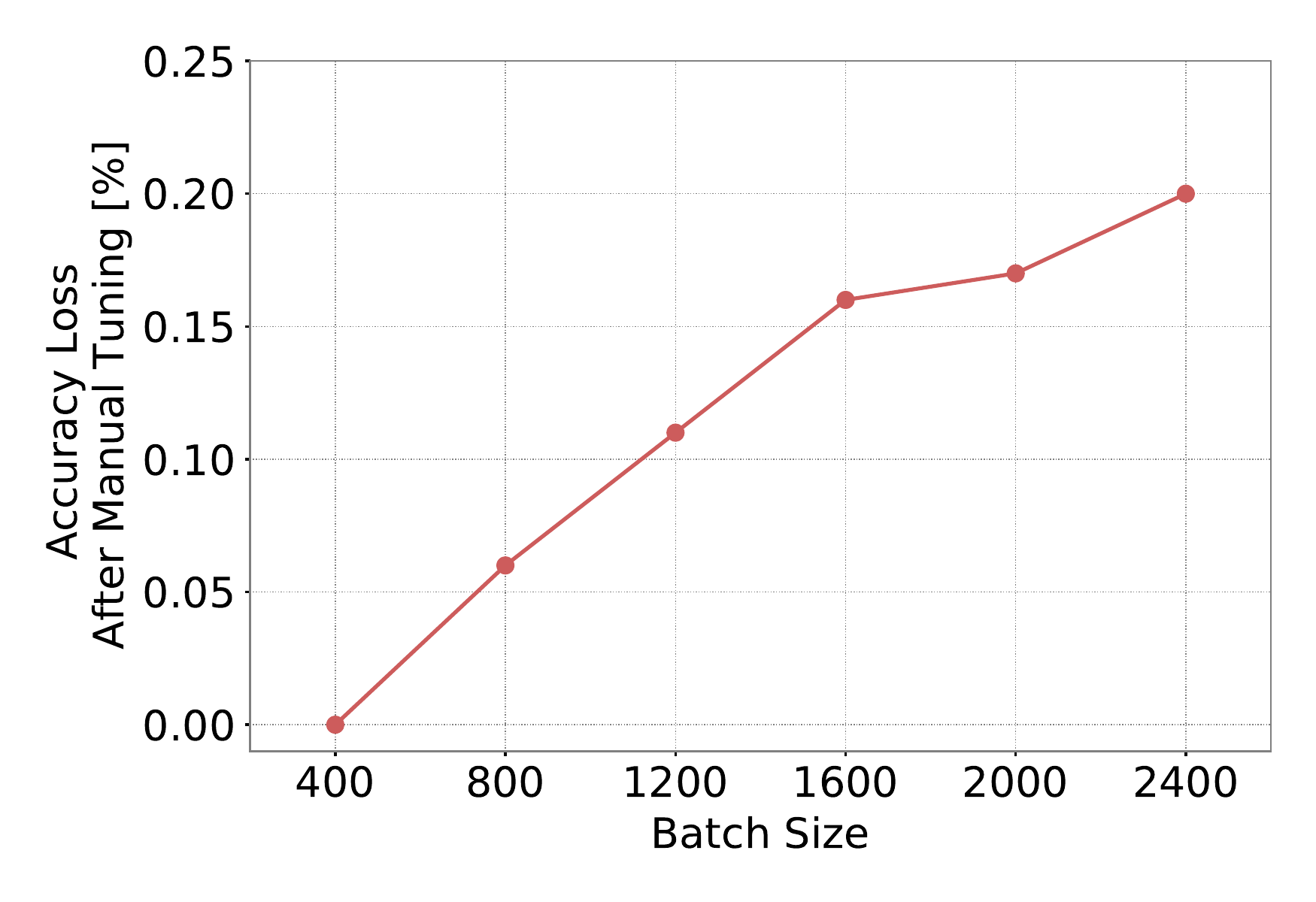}
\vspace{-0.2cm}
\caption{Accuracy gap on GPUs compared with CPU runs increase with the batch size when the model is tuned manually.}
\vspace{-0.3cm}
\label{accuracy_vs_batch}
\end{figure}

Different model and system configurations result in different optimal hyper-parameters, impacting the model quality, measured by the convergence of traditional model loss metrics, such as normalized entropy or mean squared error. To achieve optimal model convergence, hyper-parameters such as the learning rate needs
to be re-tuned as the batch size and the number of
servers change~\cite{goyal2017accurate}.
Figure~\ref{accuracy_vs_batch} shows the model loss on GPU
as the batch size is scaled after manual hyper-parameter tuning.
Despite the tuning, accuracy gap grows as we scale the batch size.
Model loss regression
in the order of 0.1-0.2\% might be considered small in
certain machine learning use cases, however for recommendation models,
such model accuracy trade-off may not be tolerable.
This makes an automated approach for hyper-parameter tuning necessary to achieve a similar or higher model quality.

FBLearner, a suite of ML tools at Facebook~\cite{dunnintroducing}, supports a parameter sweep feature with different search strategies
such as grid-based, random, and Bayesian optimization.
Users can select a search strategy to find optimal parameters automatically -- a process often known as AutoML.
We use a Bayesian optimization based strategy~\cite{bakshy2018ae} to re-tune the hyper-parameters in our GPU setup from scratch. 
For both M1\textsubscript{prod} and M2\textsubscript{prod}
GPU setup showed higher accuracy, i.e. lower negative entropy (NE) (-0.2\%, -0.1\% respectively).
Using less number of trainers and higher synchronization
rate are possible reasons for achieving higher model quality.

The process of hyper-parameter tuning to minimize model loss took around a week
to complete as we used high volumes of data to train to
ensure the quality of the new model setup.
Due the model's low throughput and the cost of running
a hyper-parameter search sweep, accuracy was not evaluated for M3\textsubscript{prod}.

\section{Related Work}
\label{related}
Using accelerators for recommendation models have started to gain
momentum in the industry. 
For example, as compared to CPUs, Google's search and ranking recommendation model training 
performs 14 times better on TPUs~\cite{dean2017machine}. 
Baidu showed a hierarchical GPU parameter server approach
to train a 10 TB recommendation model that is two times faster than CPU
training~\cite{zhao2020distributed}.
Alibaba’s advertisement recommendation models use GPUs, 
at least, for inference~\cite{zhou2019deep}. 

Most recently, in 2020, Google scored the first place in the 
MLPerf benchmark competition~\cite{mattson2020mlperf} for 
training Facebook's open source Deep Learning Recommendation Model (DLRM)~\cite{naumov2019deep}
on TPUv4 in just 1.21 minutes~\cite{google_mlperf}.
Furthermore, NVIDIA's DGX-A100 with its own Merlin HugeCTR software stack 
won the second place, finishing DLRM training in 3.33 minutes~\cite{nvidia_mlperf}.
NVIDIA's optimized software library enables more than 30\% performance 
speedup for the DGX-A100 systems (and more than 70\% speedup for DGX2-V100).

Despite the importance of deep learning recommendation models, 
this class of deep learning workloads is under studied, especially by the system and architecture's community~\cite{wu:sigarch2019}. 
Thus far, industry-scale recommendation models are represented in the MLPerf Training and Inference benchmark suites~\cite{mattson2020mlperf,mlperf-inference,mlperf-reco-advisory,mlperf-training}. However, the MLPerf-DLRM benchmark represents 
a medium-scale recommendation model for click-through-rate prediction.
As what we presented in this paper, there is a diverse 
collection of recommendation models deployed at Facebook's production environment.
Depending on the specific recommendation use cases, model architectures and parameters used for training can vary. 
In particular, efficient training for the large embedding tables with varying memory access patterns imposes significant 
system design and optimization challenges. 
In addition, recent studies have started analyzing the system- and architecture-level implications of neural recommendation inference~\cite{gupta2020deeprecsys, gupta2020architectural,hsia:iiswc20}. Recent works also examine near memory processing architectures, such as RecNMP~\cite{ke2020recnmp}, TensorDIMM~\cite{kwon2019tensordimm}. However, neither RecNMP nor TensorDIMM is optimized for gradient aggregation. The benefits do not translate well into training performance improvement.
Finally, making training infrastructures reliable has a profound impact in the training workflow efficiency as well~\cite{scar,maeng2020cpr,nicolae:hal-02543977}.

This paper is the first 
to describe the unique properties of industry-scale recommendation 
models trained at Facebook's production environment. 
The key characteristics of the model architecture with MLP stacks and a collection of large embedding tables underpin the design of Facebook's next-generation Zion systems~\cite{zion}. 
The goal of this paper is to advance the state-of-the-art understanding of
Facebook's deep learning recommendation models, the wide variety of model parameters and architectures and the complexity of the hardware-software co-design space for Facebook's production scale. 
We also present in-depth training throughput and model quality characterization analysis 
for three production-representative models.
We describe the optimization techniques necessary to enable recommendation training on Facebook's existing Big Basin and next-generation Zion 
training systems.


\section{Conclusion}
\label{conclusion}

Deep learning recommendation models have diverse set of characteristics
based on their model architectures and parameter configurations, leading to different levels of 
CPUs, memory capacity, memory and network bandwidth requirements. 
We share insights from Facebook's production workflows by characterizing 
the effects of dense and sparse features, batch sizes, 
embedding table hash sizes, and MLP dimensions, on training throughput.

Important challenges continue to loom for training deep learning
recommendation models as the memory capacity requirement for embedding tables grow into multiple
terabytes. Furthermore, training requires high volumes of data to ensure
the model quality.
The need for high throughput coupled with high accuracy
is driving the industry to design and customize specialized
hardware architectures for training.
We present the next-generation scale-out Zion platform to address this need.
The design space characterization and analysis presented in this 
paper with Facebook's production-scale deep learning recommendation models can be used to guide the design of training infrastructures.
We hope the insights will enable further research across the entire system stack, from the design of training algorithms to next-generation training infrastructure development and optimization. 


\section{Acknowledgements}
We would like to thank Facebook colleagues, especially Shunting Zhang, Hassan Eslami, Chenguang Xi, Manoj Krishnan, Jiyan Yang for the discussions and feedback on this work.



\bibliographystyle{IEEEtranS}
\bibliography{ref}

\end{document}